\documentclass[preprint,11pt,3p]{elsarticle}

\usepackage{subcaption}
\usepackage{amsmath}
\usepackage{hyperref}
\usepackage{amssymb}
\hypersetup{colorlinks,linkcolor={blue},citecolor={blue},urlcolor={red}} \usepackage{nicematrix} 
\usepackage{floatrow}
\usepackage{colortbl}
\usepackage{booktabs}  
\usepackage[export]{adjustbox}
\usepackage{cases}
\usepackage{verbatim}
\usepackage[title]{appendix}
\usepackage{tikz}
\usepackage[mathcal]{euscript}
\usepackage[final]{pdfpages}
\usepackage{pdfpages}
\usepackage{diagbox}
\usepackage[integrals]{wasysym}
\usepackage{amsfonts} 
\usepackage{varioref} 
\usepackage{float}
\usepackage{epsfig}
\usepackage{epstopdf}
\usetikzlibrary{shadings}
\usetikzlibrary{arrows}
\usetikzlibrary{shapes}
\usetikzlibrary{decorations}
\usepackage{bm}
\usepackage{mathtools}
\usepackage{graphicx}
\usepackage{caption}
\usepackage{tabularx}
\renewcommand{\arraystretch}{1.5}
\usepackage{graphicx}
\usepackage{subcaption}
\usepackage{booktabs}
\usepackage{multirow}

\usepackage{breqn}
\usepackage{array}
\usepackage{stmaryrd}
\usepackage{amsmath}
\usepackage{mathtools}

\DeclarePairedDelimiter\abs{\lvert}{\rvert}%
\DeclarePairedDelimiter\norm{\lVert}{\rVert}%

\makeatletter
\let\oldabs\abs
\def\abs{\@ifstar{\oldabs}{\oldabs*}}
\let\oldnorm\norm
\def\norm{\@ifstar{\oldnorm}{\oldnorm*}}
\makeatother
\usepackage{algorithm}
\usepackage{algpseudocode}
\usepackage{lipsum,graphicx}
 \usepackage{lineno}

\begin{document}

%\begin{linenumbers} 
\begin{frontmatter}
 \title{Physics-informed neural networks   for solving moving interface flow
problems using the level set approach} 
\author{Mathieu Mullins $^{a}$, Hamza Kamil$^{a,b}$, Adil Fahsi$^{c}$, Azzeddine Soulaïmani$^{a}$}
 \address{$^a$  École de technologie supérieure, Canada}
 \address{$^b$   University Mohammed VI Polytechnic, Morocco}
 \address{$^c$   University Sultan Moulay Slimane, Morocco}

\cortext[mycorrespondingauthor]{Corresponding author: Azzeddine Soulaïmani \newline
Email address: azzeddine.soulaimani@etsmtl.ca}

\begin{abstract}
%\begin{linenumbers} 

This paper advances the use of physics-informed neural networks (PINNs) architectures to address moving interface problems via the level set method. Originally developed for other PDE-based problems, we particularly leverage PirateNet’s features—including causal training, sequence-to-sequence learning, random weight factorization, and Fourier feature embeddings—and tailor them to handle problems with complex interface dynamics. Numerical experiments validate this framework on benchmark problems such as Zalesak’s disk rotation and time-reversed vortex flow. We demonstrate that PINNs can efficiently solve level set problems exhibiting significant interface deformation without the need for upwind numerical stabilization, as generally required by classic discretization methods, or additional mass conservation schemes. However, incorporating an Eikonal regularization term in the loss function with an appropriate weight can further enhance results in specific scenarios. Our results indicate that PINNs with the PirateNet architecture surpass conventional PINNs in accuracy, achieving state-of-the-art error rates of \(L^2=0.14\%\) for Zalesak's disk and \(L^2=0.85 \%\) for the time-reversed vortex flow problem, as compared to reference solutions. Additionally, for a complex two-phase flow dam break problem coupling the level set with the Navier-Stokes equations, we propose a geometric reinitialization method embedded within the sequence-to-sequence training scheme to ensure long-term stability and accurate inference of the level set field.
%\end{linenumbers} 
\end{abstract}


\end{frontmatter}

\setcounter{tocdepth}{3}  
%\end{linenumbers} 
%\linenumbers 

\section{Introduction}
\label{intro}
Studying moving interface problems, such as the interaction and evolution of distinct phases or boundaries, is fundamental to various engineering disciplines, including fluid mechanics, materials science, and medical imaging. Modeling free surfaces and multiphase flow phenomena is essential for a wide range of industrial and natural applications, as they support the development of efficient and safe products, such as ships, oil pipelines, and water treatment systems, as well as policies for environmental protection and disaster management \cite{cheng_improved_2024, google_research_flood_2023}. 

Traditionally, computational fluid dynamics (CFD) has provided two primary approaches for numerically solving multiphase flow problems: Lagrangian and Eulerian methods. Lagrangian methods track the interface using a mesh that moves and adapts with the flow, which allows for highly accurate simulations of the interface evolution. However, this approach becomes challenging to implement in three-dimensional (3D) problems, particularly when dealing with complex or highly deformable interfaces. In contrast, Eulerian methods, including techniques such as front-tracking and front-capturing, employ a fixed computational mesh that remains stationary throughout the simulation. Eulerian approaches are more adaptable to complex geometries and large deformations. However, these approaches require fine mesh resolution and significant changes in the discretization formulations to ensure accuracy near the interface. The following discusses Eulerian methods.

Front-tracking methods use markers to track the evolution of the flow and are divided into volume-tracking and surface-tracking techniques. From the volume-tracking family, the Marker and cell method has been used to solve the dam-break problem \cite{harlow_numerical_1965}. However, Harlow and Welch noted that the method required frequent redistribution of markers to preserve accuracy \cite{harlow_numerical_1965}. Surface-tracking is usually more accurate than the volume-tracking counterpart. Among other applications, some surface tracking methods have been developed to solve the rising bubble and merging bubble problems \cite{unverdi_front-tracking_1992}.

On the other hand, front-capturing methods follow the evolution of a scalar function related to the position of the interface. Their dependence on a simple scalar function makes the main methods of Volume of Fluid (VOF) and Level Set easy to implement. In VOF, the function used is the volume fraction of a phase and has been used to solve complex fluid-surface interaction problems in \cite{noh_slic_1976} using the simple line interface calculation.
In contrast, the level set method defines the moving interface as the zero contour of a higher-dimensional signed distance function (SDF), enabling the interface to change shape and topology seamlessly. Initially developed by \cite{osher_fronts_1988} to simulate flow problems, the authors showed that the level set method could accurately capture the formation of sharp gradients and cusps in moving fronts, making them great at handling problems with breaking and complex topologies. Since the method does not require the moving surface to be written as a function, it can be applied to more general Hamilton-Jacobi-type problems \cite{osher_fronts_1988}. This has been done repeatedly in different fields, such as shape modeling in computer vision \cite{malladi_shape_1995} and topology optimization \cite{wang_level_2003}.

The standard level set method faces some challenges with the accumulation of errors. Being first defined by the SDF, a transport equation is then solved to allow the motion of the interface to be driven by a given velocity field. The accumulation of errors in the transport equation can cause an important loss of smoothness at the interface and a loss of mass, depending on the simulation parameters. Hence, a lot of research has focused on developing reinitialization algorithms to mitigate these issues. Nonetheless, there remain difficulties with reinitialization, notably when choosing the delay between reinitializations and the chosen algorithm \cite{gomes_reconciling_2000}. Some techniques have been developed to eliminate the need for reinitialization, but these can increase the computational cost \cite{toure_stabilized_2016}.

With the rapid development of scientific machine learning in recent years, researchers have explored new ways to solve time-dependent problems. Among promising methods, Physics-Informed Neural Networks (PINN) have emerged as a compelling approach to modeling complex systems by integrating physical laws directly into the training objective. 

Introduced by Raissi et al. \cite{raissi_physics-informed_2019}, PINNs solve forward, inverse, and mixed problems governed by partial differential equations (PDE). Physical laws are integrated into the training process through a regularization term added to the loss function. This regularization is obtained by applying a differential operator directly to the network outputs. The resulting expression is then evaluated and minimized across selected collocation points within the computational domain. The problem is essentially converted from trying to solve the governing equations to a loss function optimization problem. This process is possible due to automatic differentiation (AD) \cite{baydin_automatic_2017}, which uses the chain rule to calculate the derivatives of mathematical expressions found in neural networks (NN). Furthermore, PINNs are mesh-free and can learn completely unsupervised, thus removing the need for expensive labeled training data. This makes them attractive for solving data-scarce of high-dimensional problems.

Since their inception, PINNs have achieved great results across various fields of science and engineering like materials science \cite{fang_deep_2020,goswami_physics-informed_2022,liu_physics-informed_2021,salvati_defect-based_2022}, energy \cite{cho_lstm-pinn_2022,pombo_increasing_2022}, thermodynamics \cite{liu_surrogate_2023,masclans_thermodynamics-informed_2023} and fluid dynamics \cite{dazzi_physics-informed_2024,donnelly_physics-informed_2024,li_improved_2024,qi_physics-informed_2024,kamil_transfer_2024}. Despite their success in solving benchmark problems, vanilla PINNs as implemented by Raissi et al. reveal some training disorders such as spectral bias \cite{rahaman_spectral_2019,wang_eigenvector_2021},  unbalanced gradient flow   \cite{wang_understanding_2021,wang_when_2022} and violation of temporal causality \cite{wang_respecting_2022}. Several studies have proposed strategies to address these challenges by modifying the deep learning architecture. Notable approaches include the introduction of a modified Multi-Layer Perceptron (MLP) architecture \cite{wang_understanding_2021}, the use of random Fourier features for improved function representation \cite{rahaman_spectral_2019, tancik_fourier_2020}, random weight factorization for enhanced training dynamics \cite{wang_random_2022}, and adaptive resampling of collocation points to better capture important regions in the solution domain \cite{daw_mitigating_2023, wu_comprehensive_2023}. Other techniques applied to the training algorithm can also improve PINNs’ abilities, such as respecting temporal causality \cite{wang_respecting_2022}, automatic re-weighting of loss terms \cite{chen_gradnorm_2018,wang_understanding_2021,wang_when_2022}, curriculum training \cite{bengio_curriculum_2009,krishnapriyan_characterizing_2021} and Sequence-to-Sequence (S2S) learning \cite{krishnapriyan_characterizing_2021}. These improvements have been resumed by Wang et al. in \cite{wang_experts_2023} along with their release of a highly optimized JAX library tailored to developing PINN architectures meant to solve PDE-based problems. They recently extended their Jaxpi library with the \textit{PirateNet} architecture allowing the use of deeper networks along with physics-informed initialization to obtain better results \cite{wang_piratenets_2025}.

The literature suggests that Physics-Informed Neural Networks (PINNs) can successfully solve a variety of PDE-driven problems; however, little research has been devoted to applying PINNs to solve the level set equation, especially in complex fluid problems. In \cite{m_silva_pinn-based_2024} and \cite{zhou_self-adaptive_2024}, a PINN-based level set was used to solve a rising bubble problem in a constant flow field. More recently, \cite{tang_physics-informed_2024} used a level set PINN to solve a vapor condensation problem. To the best of the authors' knowledge, this is the first study to solve a level set problem using a PINN-based model in a complex varying velocity field that demonstrates strong vorticity and stretching of the interface.

This study introduces a PINN-based framework for solving level set equations in moving interface problems and extends it to coupled level set–Navier-Stokes systems with geometric reinitialization for long-term stability. Our contributions include leveraging advanced features of the \textit{PirateNet} architecture—such as causal training, sequence-to-sequence learning, and random weight factorization—and tailoring them to handle problems with complex interface dynamics.

Our study begins by testing our PINN approach on level set problems, starting with a fabricated simple uniform transport case to highlight the challenges faced by the standard PINN solver and demonstrate the advantages of the proposed techniques. Next, we tackle two well-known level set problems: the rigid-body rotation of Zalesak's disk \cite{zalesak_fully_1979} and a more complex time-reversed vortex flow problem, which features a varying velocity field \cite{toure_stabilized_2016}. For each case, we compare the reference solutions to results obtained from (a) the original PINN formulation \cite{raissi_physics-informed_2019}, (b) the improved PINN approach \cite{wang_experts_2023}, and (c) the state-of-the-art \textit{PirateNet} framework \cite{wang_piratenets_2025}.  We finish with a challenging two-phase flow dam break application which couples the level set to the incompressible Navier-Stokes equations. Our findings demonstrate that \textit{PirateNet} can solve complex level set problems effectively without requiring reinitialization or mass conservation schemes for an exactly conservative flow field. However, for more challenging problems where training can be more unstable, geometric reinitialization can be used to enable long term inference of the level set field.

Our main contributions are summarized as follows:
\begin{enumerate}
    \item We conduct various numerical experiments showing that \textit{PirateNets}' architecture outperforms both the original and improved PINN models in level set benchmark problems, especially those involving significant interface deformation. 
    \item With experiments on classical level set benchmarks, we show that PINNs can learn the level set equations without the need for upwind numerical stabilization or mass conservation schemes.
    \item We show  that adding an Eikonal regularization term to the loss function with an appropriate weight can improve the results in certain cases.
    \item For complex level set problems were the signed distance property is lost through time, we propose a geometric reinitialization method embedded in the sequence-to-sequence training scheme that allows long term inference of the level set field.
\end{enumerate}

The rest of the paper is organized as follows: In Section \ref{levelset_method}, the level set method is described along with classic reinitialization schemes.  In Section \ref{physics-informed_nn}, we review the PINN, improved PINN and \textit{PirateNet} architectures. Section \ref{results} presents our numerical tests followed by the main conclusions in Section \ref{Conclusion}.

\section{The level set method}
\label{levelset_method}
In the following, we present the level set approach as a framework for tracking interfaces in multiphase flow problems. 

\subsection{General level set formulation}
The level set method, introduced by Osher and Sethian \cite{osher_fronts_1988}, is an efficient tool for tracking interface motion in multiphase flow problems. This method utilizes a scalar level set function, \(\phi(\bm{x}, t)\), where the zero-level set \(\phi(\cdot, t) = 0\) defines the interface separating two or more phases. Mathematically, the interface at time $t$ is represented as:
\begin{equation}
    \Gamma(t) = \biggl\{\bm{x} \in \mathbb{R}^d \;\big|\; \phi(\bm{x}, t) = 0\biggr\},
    \label{eq:ls_interface}
\end{equation}
where \(d = 2, 3\) denotes the spatial dimensions.

\begin{figure}
    \centering
    \includegraphics[width=0.95\linewidth]{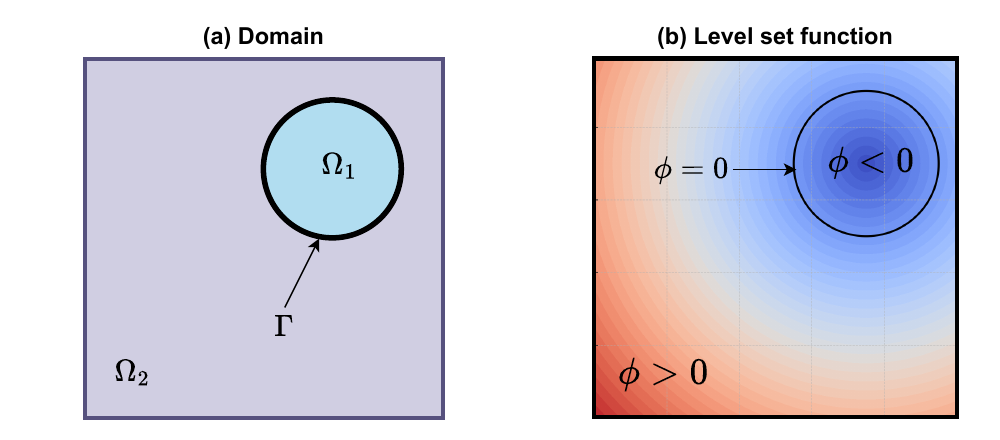}
    \caption{Illustration of a two-phase domain (a), with \(\Omega_i\) representing phase \(i\). (b) Depicts an example of a level set function corresponding to the domain in (a).}
    \label{fig:level_set_domain}
\end{figure}

Figure \ref{fig:level_set_domain} provides an example of a two-phase flow problem. In this example, \(\Omega_i(t)\) represents the region occupied by phase \(i\) at time \(t\), with the entire domain given by \(\Omega = \Omega_1 \cup \Omega_2\). The interface separating the subdomains, \(\Omega_1\) and \(\Omega_2\), is \(\Gamma(t) = \Omega_1 \cap \Omega_2\). The level set function \(\phi\) is defined with a sign convention as follows:
\begin{equation}
    \begin{cases}
       \phi(\bm{x}, t) < 0 & \text{if } \bm{x} \in \Omega_1,\\
     \phi(\bm{x}, t)   = 0 & \text{if } \bm{x} \in \Gamma,\\
     \phi(\bm{x}, t)   > 0 & \text{if } \bm{x} \in \Omega_2.
    \end{cases}
    \label{eq:ls_interface_sign}
\end{equation}

The level set function \(\phi\) is commonly defined as a signed distance function (SDF) $d_{\text{signed}}$, which is expressed as:
\begin{equation}
    d_{\text{signed}}(\bm{x}, t) =
    \begin{cases}
        - d(\bm{x}, t) & \text{if } \bm{x} \in \Omega_1,\\
        0 & \text{if } \bm{x} \in \Gamma,\\
        + d(\bm{x}, t) & \text{if } \bm{x} \in \Omega_2,
    \end{cases}
    \label{eq:ls_sdf1}
\end{equation}
where \(d(\bm{x}, t)\) is a distance function from the interface $\Gamma$:
\begin{equation}
    d(\bm{x}, t) = \min_{\bm{x}_\Gamma \in \Gamma} \|\bm{x} - \bm{x}_\Gamma\|.  
    \label{eq:ls_sdf2}
\end{equation}

The SDF $d_{\text{signed}}$ satisfies the following Eikonal equation:
\begin{equation}
    ||\nabla d_{\text{signed}}||=1.
    \label{eq:ls_grad1}
\end{equation}

The level set approach tracks the evolution of the interface \(\Gamma(t)\) over time by identifying the zero level set of the solution to the transport equation:   
\begin{equation}
    \frac{\partial \phi}{\partial t} + \bm{u} \cdot \nabla \phi = 0,
    \label{eq:ls_eqn}
\end{equation}  
where \(\bm{u}\) represents the velocity field. Initially, the level set function \(\phi\) is defined as the SDF \(\phi(\bm{x}, 0) = d_{\text{signed}}(\bm{x}, 0)\).

\subsection{Reinitialization}

Traditional numerical methods for solving the level set equation often suffer from error accumulation, which can cause the level set function to lose its signed distance property. Preserving this property is crucial for accurately computing geometric quantities such as curvature and interface normal vectors, which are essential for many applications.

To address this issue, reinitialization methods have been introduced to restore the level set function to its signed distance property.
The main idea is to ensure that the gradient of the level set function remains close to 1, \(||\nabla \phi||\simeq 1\).

One such method, proposed by Sussman, Smereka, and Osher \cite{sussman_level_1994}, iteratively adjusts the level set function by solving a PDE over a pseudo-time.

Some articles have explored different reinitialization methods, such as the geometric reinitialization \cite{fahsi_numerical_2017,ausas_geometric_2011} which reinitializes the level set function by measuring the distance from the nodes to the level set interface and the Fast Marching method \cite{sethian_fast_1996,sethian_theory_1996}. Others have also found ways to bypass the reinitialization step using variational energy methods \cite{li_level_2005,toure_stabilized_2016}.

\subsection{Mass conservation}

Mass conservation in level set methods ensures that the total area of each phase or material represented by the level set function remains constant over time. This property is important, as numerical errors can lead to artificial loss or gain of mass, particularly in coarser grids or during long-term simulations. Maintaining mass conservation is critical when using level set approaches, as it is not inherently guaranteed at the discrete level.

Despite various efforts to improve mass conservation \cite{van_der_pijl_mass-conserving_2005,olsson_conservative_2005,olsson_conservative_2007}, achieving highly accurate mass conservation remains a significant challenge.

For a conservative flow field, the total area of each phase represented by the level set function should remain constant over time, as described by the global area conservation equations:

\begin{equation}
    \frac{dA_1}{dt} = 0 \quad \text{for } \phi(\bm{x}, t) < 0,
    \quad \frac{dA_2}{dt} = 0 \quad \text{for } \phi(\bm{x}, t) > 0,
    \label{eq:ls_area_conservation}
\end{equation}
\noindent 
where \(A_1\) and \(A_2\) represent the areas of the two phases separated by the zero level set. These equations express the requirement that the enclosed areas remain invariant over time, ensuring mass conservation in the continuous formulation.

\section{Methodology}
\label{physics-informed_nn}
In the following section, we build on the previous discussion to provide the adopted strategy for solving complex level set problems.
\subsection{Physics-informed neural networks (PINNs)}
\label{pinn}
Consider a general PDE of the form:
\begin{equation}
    \bm{u}_t(\bm{x}, t) + \mathcal{N}[\bm{u}](\bm{x}, t) = 0, \quad t \in (0, T], \ \bm{x} \in \Omega,
    \label{eq:general_pde}
\end{equation}
where \(\bm{u}\) is the unknown solution of the PDE, \(\mathcal{N}[\cdot]\) represents a linear or nonlinear differential operator, \(t\) denotes the temporal coordinate, \(\bm{x}\) represents the spatial coordinates, \(\bm{u}_t\) denotes the partial derivative of \(\bm{u}\) with respect to time, \(\Omega\) is a subset of \(\mathbb{R}^D\) defining the spatial domain, and \(T\) is the final simulation time. The PDE~\eqref{eq:general_pde} is subject to the following initial and boundary conditions:
\begin{equation}
    \bm{u}(\bm{x},0) = g(\bm{x}), \quad \bm{x} \in \Omega,
    \label{eq:ic_pde}
\end{equation}
\begin{equation}
    \mathcal{B}[\bm{u}] = 0, \quad t \in (0, T], \quad \bm{x} \in \partial\Omega,
    \label{eq:bc_pde}
\end{equation}
\noindent
where $g$ is the initial condition, and \( \mathcal{B}[\cdot]\) is a boundary operator (Dirichlet, Neumann, Robin, periodic).
We approximate the solution of PDE~\eqref{eq:general_pde} using a deep neural network \(\bm{u}_\theta(t,\bm{x})\), where \(\theta\) represents the set of trainable parameters. The PINN framework finds the solution by minimizing a composite loss function:
\begin{equation}
    \mathcal{L}(\theta) = \lambda_{ic}\mathcal{L}_{ic}(\theta) + \lambda_{bc} \mathcal{L}_{bc}(\theta) + \lambda_{r}\mathcal{L}_r(\theta),
    \label{eq:loss_function}
\end{equation}
where each component enforces a specific constraint of the problem. The term \(\mathcal{L}_{ic}(\theta)\) corresponds to the initial condition, \(\mathcal{L}_{bc}(\theta)\) accounts for the boundary condition, and \(\mathcal{L}_r(\theta)\) ensures that the physical laws governing the system are satisfied. The weights \(\lambda_{ic}\), \(\lambda_{bc}\), and \(\lambda_{r}\) balance the influence of these constraints in the overall loss function, directing the network to learn a solution that adheres to all problem requirements.

The initial condition loss is given by:

\begin{equation}
    \mathcal{L}_{ic}(\theta) = \frac{1}{N_{ic}} \sum_{i=1}^{N_{ic}} |\bm{u}_\theta( x_{ic}^i,0) - g(x_{ic}^i)|^2,
    \label{eq:loss_ic}
\end{equation}
 The boundary condition loss is defined as:
\begin{equation}
    \mathcal{L}_{bc}(\theta) = \frac{1}{N_{bc}} \sum_{i=1}^{N_{bc}} |\mathcal{B}[\bm{u}_\theta](x_{bc}^i, t_{bc}^i)|^2,
    \label{eq:loss_bc}
\end{equation}
 The residual loss ensures adherence to the PDE and is defined as:
\begin{equation}
    \mathcal{L}_r(\theta) = \frac{1}{N_r} \sum_{i=1}^{N_r} |\mathcal{R}_\theta(x_r^i, t_r^i)|^2,
    \label{eq:loss_res}
\end{equation}
where the residual \(\mathcal{R}_\theta(x, t)\) is given by:
\begin{equation}
    \mathcal{R}_\theta(x, t) = \frac{\partial \bm{u}_\theta}{\partial t}(x_r, t_r) + \mathcal{N}[\bm{u}_\theta](x_r, t_r).
    \label{eq:residual}
\end{equation}
\noindent
 The terms \(\{x_{ic}^i\}_{i=1}^{N_{ic}}\), \(\{t_{bc}^i, x_{bc}^i\}_{i=1}^{N_{bc}}\), and \(\{t_r^i, x_r^i\}_{i=1}^{N_r}\) represent sets of collocation points sampled within the computational domain to evaluate the initial condition, boundary condition, and residual losses, respectively.
Automatic differentiation is used to compute the required partial derivatives, enabling efficient evaluation of \(\mathcal{R}_\theta(t, x)\). This approach ensures that the neural network predicts solutions that satisfy the initial and boundary conditions as well as the governing PDE.

Generally, the standard PINN solver encounters several challenges when applied to certain types of problems. These challenges include temporal causality violation \cite{wang_respecting_2022}, difficulties handling long-time simulations \cite{penwarden_unified_2023}, and various other failure modes. As mentioned in the introduction, these limitations necessitate enhancements to the vanilla PINN architecture.
Several improvements have been proposed to address these issues and improve the robustness and efficiency of PINNs. In the following, we present the modified PINN solver adopted in our study, which incorporates some of these advancements.

\subsection{Improved PINNs}
\label{sec:Improved_PINNs}
This section discusses the main improvements over the vanilla PINN.

\subsubsection{PirateNet architecture}
\label{piratenet_arch}
\begin{figure}
    \centering
    \includegraphics[width=0.9\linewidth]{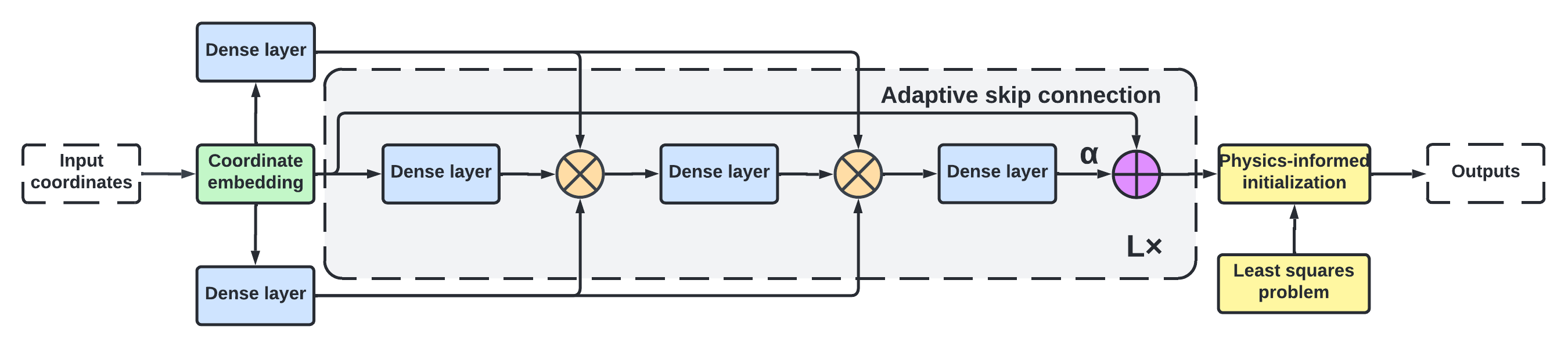}
    \caption{\textit{PirateNets'} architecture \cite{wang_piratenets_2025}. The input coordinates are passed to a coordinate embedding before going through two dense layers and the residual block. The residual block is contained within the grey area in the figure and is repeated \(L\) times. The initial dense layers are passed to two gating operations in orange within the residual block. The adaptive skip connection is affected by the trainable parameter \(\alpha\) and controls how much information is passed directly from the coordinate embedding. Finally, a physics-informed initialization scheme is applied before the output.}
    \label{fig:PirateNet_architecture}
\end{figure}

In most PINN studies, the Multi-Layer Perceptron (MLP) is commonly used as the deep learning architecture. MLPs are usually effective for simple problems, but improvements have been proposed to enhance their capabilities. For example, Wang et al. proposed a modified MLP \cite{wang_understanding_2021} that introduces two encoders to the input coordinates, which generally improve the results with PINN. However, as MLPs deepen to handle complex solutions, they often face significant training inefficiencies. To address this challenge, Wang et al. \cite{wang_piratenets_2025} introduced \textit{PirateNets}, which leverage adaptive residual connections to improve training performance. This approach mitigates training inefficiencies by starting with an artificially shallow network and progressively increasing its depth during training. The adaptive skip connections within the residual block enable efficient optimization and improve training stability. A schematic of the \textit{PirateNet} forward pass is shown in Figure~\ref{fig:PirateNet_architecture} and is described in detail below.

The input coordinates \(\bm{x} = (x, t)\) are first transformed into a higher-dimensional feature space using a coordinate embedding technique, such as random Fourier features \cite{tancik_fourier_2020}:
\begin{equation}
\Phi(\bm{x}) = \left[\cos(\textbf{F}\bm{x}), \sin(\textbf{F}\bm{x})\right] ^T,
    \label{eq:random_fourier_feat}
\end{equation}
\noindent
where each element of the matrix \(\textbf{F}\) is sampled from a Gaussian distribution \(\mathcal{N}(0, \sigma_F^2)\), and \(\sigma_F\) is a hyperparameter. Random Fourier features are particularly effective in reducing spectral bias, enabling improved predictions for high-frequency components in the solution \cite{wang_eigenvector_2021}.

The embedded coordinates \(\Phi(\bm{x})\) are then sent to two dense layers acting as gates within each residual block of the model:
\begin{equation}
    U=\sigma(\bm{W}_1\Phi (\bm{x})+\bm{b}_1),\qquad V=\sigma(\bm{W}_2\Phi (\bm{x})+\bm{b}_2),
    \label{eq:pirate_dense}
\end{equation}
where \(\sigma\) is a point-wise activation function.
\noindent
Let \(\bm{x}^{(l)}\) be the input of the \(l\)-th residual block where \(1\leq l \leq L - 1\). The forward pass of a \textit{PirateNet's} residual block can now be defined as follows:
\begin{equation}
\begin{aligned}
    f^{(l)}    & = \sigma (\bm{W}_1^{(l)}\bm{x}^{(l)}+\bm{b}_1^{(l)}), \quad &\text{(Dense layer)} \\
    z_1^{(l)}  & = f^{(l)}\odot U+(1-f^{(l)}) \odot V, \quad &\text{(Gating operation)} \\
    g^{(l)}    & = \sigma (\bm{W}_2^{(l)}z_1^{(l)}+\bm{b}_2^{(l)}), \quad &\text{(Dense layer)} \\
    z_2^{(l)}  & = g^{(l)}\odot U+(1-g^{(l)}) \odot V, \quad &\text{(Gating operation)} \\
    h^{(l)}    & = \sigma (\bm{W}_3^{(l)}z_2^{(l)}+\bm{b}_3^{(l)}), \quad &\text{(Dense layer)} \\
    \bm{x}^{(l+1)}    & = \alpha^{(l)}h^{(l)} + (1-\alpha^{(l)})\bm{x}^{(l)}, \quad &\text{(Adaptive skip connection)}
\end{aligned}
    \label{eq:pirate_residual_block}
\end{equation}
\noindent
where \(\odot\) is a Hadamard product and \(\alpha^{(l)}\) are trainable parameters for each residual block. All weights are initialized with the Glorot scheme \cite{glorot_understanding_2010} while the biases, \(b\), are initialized to zero. The output of a PirateNet with \(L\) blocks is finally given by $\bm{W}^{(L+1)}\bm{x}^{(L)}$.

A key component of \textit{PirateNets} is the adaptive skip connection, characterized by its trainable parameter \(\alpha^{(l)}\), which controls the degree of nonlinearity introduced by the \(l\)-th block. Initially, \(\alpha^{(l)}\) is set to zero, effectively reducing the block to an identity mapping. During training, \(\alpha^{(l)}\) gradually increases, allowing the network to dynamically introduce non-linear transformations only when they begin to meaningfully contribute to the solution. This progressive activation strategy mitigates the challenges associated with initializing deeper networks, enhancing training stability and efficiency \cite{wang_piratenets_2025}.

Another crucial component of the \textit{PirateNet} architecture is the \textit{physics-informed initialization} approach, designed to provide an optimal starting point for training by incorporating available data \cite{wang_piratenets_2025}. As mentioned before, the \textit{PirateNet} model is represented as a linear combination of Fourier basis functions at initialization, expressed as \(\bm{W}^{(L+1)}\Phi(\bm{x})\). This implies that the weight matrix \(\bm{W}^{(L+1)}\) of the final layer can be initialized by solving a least-squares problem:
\begin{equation}
   \bm{W}^{(L+1)} = \arg\min_{\bm{W}} \| \bm{W} \Phi(\bm{x}) - Y \|_2^2,
    \label{eq:pirate_least_squares}
\end{equation}
where \(Y\) represents some available data, such as initial conditions, boundary data, or outputs from surrogate models.
This initialization strategy eliminates common issues associated with random initialization, such as instability or slow convergence. By aligning the network's initial output with a physically meaningful approximation of the solution, it accelerates training and improves overall performance.

For the remainder of this work, we define the amount of layers in a \textit{PirateNet} as its amount of residual blocks. For example, a \textit{PirateNet} with 3 layers is the same as saying it has 3 residual blocks.

\subsubsection{Random weight factorization}
\label{rwf}
Random Weight Factorization (RWF) is a simple yet effective technique proposed by \cite{wang_random_2022}, which has been shown to accelerate and improve the training of neural networks. The RWF method factorizes the weights associated with each neuron in the network as follows:
\begin{equation}
   \bm{w}^{(k,l)} = \bm{s}^{(k,l)} \bm{v}^{(k,l)}, \quad k=1,2,...,d_l,\quad l=1,2,...,L+1,
    \label{eq:rwf_1}
\end{equation}
where \(\bm{w}^{(k,l)} \in \mathbb{R}^{d_l-1}\) represents the \(k\)-th row of the weight matrix \(\bm{W}^{(l)}\), \(\bm{s}^{(k,l)} \in \mathbb{R}\) is a trainable scaling factor specific to each neuron, and \(\bm{v}^{(k,l)} \in \mathbb{R}^{d_{l-1}}\).
Consequently, the RWF can be expressed in matrix form as:
\begin{equation}
   \bm{W}^{(l)} = \text{diag}(\bm{s}^{(l)}) \bm{V}^{(l)}, \quad l=1,2,...,L+1,
    \label{eq:rwf_2}
\end{equation}
where \(\bm{s}^{(l)}\in \mathbb{R}^{d_l}\) represents the vector of scaling factors for layer \(l\).

The implementation details of the RWF are as follows \cite{wang_experts_2023}. First, the neural network parameters are initialized using a standard scheme, such as the Glorot initialization \cite{glorot_understanding_2010}. Next, for each weight matrix \(\bm{W}\), a scale vector  \(\bm{s}\) is sampled from a multivariate normal distribution \(\mathcal{N}(\mu_{rwf}, \sigma_{rwf} I)\), with \(\mu_{rwf}\) and \(\sigma_{rwf}\) are hyperparameters, and \(I\) denotes the identity matrix. At initialization, the weight matrix is factorized as  $\bm{W} = \text{diag}(\exp(\bm{s})) \cdot \bm{V}$, where \(\bm{V}\) represents the remaining factorized parameter. Finally, gradient descent optimization is applied directly to the new parameters, \(\bm{s}\) and \(\bm{V}\).  
According to \cite{wang_random_2022}, a suitable selection of RWF hyperparameters can shorten the distance between the initialization and the global minimum in the factorized parameter space relative to the original parameter space within their respective loss landscapes. Furthermore, \cite{wang_experts_2023} has demonstrated through numerical experiments that, in the case of PINNs, the RWF technique enhances both accuracy and training performance across a range of PDE benchmarks.

\subsubsection{Respecting temporal causality}
\label{causal_training}
Recently, Wang et al. \cite{wang_respecting_2022} demonstrated that PINNs can violate temporal causality when solving time-dependent PDEs, potentially leading to incorrect predictions for problems that strongly depend on satisfying the initial condition (e.g., phase-field models). The authors showed that, for certain benchmarks, PINNs tend to prioritize minimizing the PDE residual at later times before accurately learning the solutions at earlier times. To address this issue, a temporal causal training approach was proposed in \cite{wang_respecting_2022}. 

The temporal causal training strategy is described as follows. First, the temporal domain \([0, T]\) is divided into \(M\) sequential segments, denoted as \(0 = t_1 < \cdots < t_M = T\). Next, the loss function \(\mathcal{L}_r(t_i, \theta)\) is defined as the PDE residual loss within the \(i\)-th temporal segment:   
\begin{equation}
    \mathcal{L}_r(t_i, \theta) = \frac{1}{N_x} \sum_{j=1}^{N_x} \left| \mathcal{R}_\theta(x_j,t_i ) \right|^2,  
\end{equation}
where \((x_j)_{j=1}^{N_x} \in \Omega\) represent points sampled from the computational domain.  

The original PDE residual loss can then be rewritten as:
\begin{equation}
   \mathcal{L}_r(\theta)=\frac{1}{M}\sum_{i=1}^{M}w_i\mathcal{L}_r(t_i, \theta),
    \label{eq:causal_1}
\end{equation}
where  \(w_i\) are weights designed to be large only when all prior residuals \(\left\{\mathcal{L}_r(t_k,\theta)\right\}_{k=1}^{i-1}\) are sufficiently minimized in order to ensure sequential temporal causality. This is achieved by defining the weights as:
\begin{equation}
   w_i=\exp\left( -\varepsilon\sum_{k=1}^{i-1} \mathcal{L}_r(t_k, \theta)\right), \quad \text{for }i=2,3,...,M,
    \label{eq:causal_2}
\end{equation}
where \(\varepsilon\) is a causality parameter that controls the steepness of the weights \(w_i\). Here, the weights \(w_i\) are inversely exponentially proportional to the cumulative residual loss from previous time steps.

The weighted residual loss can now be written as:
\begin{equation}
   \mathcal{L}_r(\theta)=\mathcal{L}_r(t_0, \theta) + \frac{1}{M}\sum_{i=2}^{M} \exp\left( -\varepsilon\sum_{k=1}^{i-1} \mathcal{L}_r(t_k, \theta)\right) \mathcal{L}_r^i(\theta),
    \label{eq:causal_3}
\end{equation}
where $\mathcal{L}_r(t_0, \theta)$ is the initial condition loss.

Equation \eqref{eq:causal_3} ensures that the PDE residual loss within a given temporal segment is not minimized until the residuals of all preceding temporal segments have been sufficiently reduced. This approach allows the solution to the PDE to evolve in a temporally causal manner, aligning with the physical propagation of information in time-dependent systems. 
The temporal causality scheme has proven effective for training stiff PDEs, particularly those characterized by strong nonlinearity \cite{wang_experts_2023, wang_respecting_2022}. We will demonstrate that integrating this approach enables the solution of complex level set problems involving large interface deformation.

\subsubsection{Loss balancing}
\label{loss_balancing}
A challenge faced when training PINNs with a composite loss function is the difference in scale of each term within the loss function during the minimization of the PDE residual. This imbalance can cause instability during stochastic gradient descent. Since the losses cannot be normalized as a pre-processing step, normalization needs to be done during training. A rudimentary approach to this problem would be to manually assign weights to each loss term before training. However, this approach is not practical, as the weights depend highly on the problem trying to be solved. The weights could be optimized using a hyper-parameter optimization scheme, but this approach is computationally expensive and sometimes impossible if no validation dataset is available. To this end, a few approaches have been developed to allow automatic rebalancing of the loss terms during training such as gradient normalization and Neural Tangent Kernel (NTK) schemes.

The gradient normalization method \cite{wang_experts_2023} involves adjusting the weight of each loss term such that the magnitudes of their gradients contribute equally to the total loss during optimization. We begin by computing the global weights \(\hat{\lambda}\) as follows:
\begin{equation}
    \begin{aligned}
        \hat{\lambda}_{ic}    & = \frac{||\nabla_{\theta}\mathcal{L}_{ic}(\theta)||+||\nabla_{\theta}\mathcal{L}_{bc}(\theta)||+||\nabla_{\theta}\mathcal{L}_{r}(\theta)||}{||\nabla_{\theta}\mathcal{L}_{ic}(\theta)||},\\
        \hat{\lambda}_{bc}    & = \frac{||\nabla_{\theta}\mathcal{L}_{ic}(\theta)||+||\nabla_{\theta}\mathcal{L}_{bc}(\theta)||+||\nabla_{\theta}\mathcal{L}_{r}(\theta)||}{||\nabla_{\theta}\mathcal{L}_{bc}(\theta)||},\\
        \hat{\lambda}_{r}    & = \frac{||\nabla_{\theta}\mathcal{L}_{ic}(\theta)||+||\nabla_{\theta}\mathcal{L}_{bc}(\theta)||+||\nabla_{\theta}\mathcal{L}_{r}(\theta)||}{||\nabla_{\theta}\mathcal{L}_{r}(\theta)||},\\
    \end{aligned}
    \label{eq:grad_norm_global_w}
\end{equation}
where \(||\cdot||\) represents the \(L^2\) norm.
If other loss terms are present, they need to be added to the global weights calculations. 
Equation \eqref{eq:grad_norm_global_w} establishes the following relation:
\begin{equation}
   ||\hat{\lambda}_{ic}\nabla_{\theta}\mathcal{L}_{ic}(\theta)|| = ||\hat{\lambda}_{bc}\nabla_{\theta}\mathcal{L}_{bc}(\theta)|| = ||\hat{\lambda}_{r}\nabla_{\theta}\mathcal{L}_{r}(\theta)||.
    \label{eq:grad_norm_avg}
\end{equation}
This ensures that the gradients of all weighted loss terms have equal magnitudes, preventing the model from prioritizing the minimization of specific terms over others during training.
The weights can then be updated using a moving average, as follows:
\begin{equation}
   \lambda_{new} = \alpha \lambda_{old} + (1-\alpha)\hat{\lambda}_{new}.
    \label{eq:grad_norm_grads}
\end{equation}
where \(\alpha=0.9\) is a hyper-parameter quantifying the influence of the old weights on the new ones. The weighting update is done at a user-specified frequency, thus making the computational overhead negligible. Unless otherwise specified, it has been set at every 500 iterations for our experiments.

The NTK method leverages the NTK matrices of PINNs associated with each loss term to determine the weights \cite{wang_when_2022,wang_experts_2023}. The NTK matrix captures the relationship between the network's parameters and its outputs, with its trace (sum of eigenvalues) indicating the convergence rate of a loss term. By computing the trace for each NTK matrix, the weights are defined to ensure comparable convergence rates across all loss terms. This method allows for more stable weight updates than with gradient normalization, but it is more computationally expensive \cite{wang_experts_2023}. Considering this, our experiments were computed using the gradient normalization method.

\subsubsection{Sequence-to-sequence training}
\label{seq2seq}
Traditional PINNs approach learning PDEs for the entire space-time, meaning they try to predict the solution for all locations at all time points. However, many physical phenomena exhibit strong sequence dependencies where the output at a given time step relies heavily on previous time steps. In these cases, sequence-to-sequence (S2S) learning \cite{krishnapriyan_characterizing_2021} could provide better results.

In the S2S approach, the temporal domain \([0, T]\) is divided into multiple time windows, \(I_i = [t_i, t_{i+1}]\), where \(t_i = i\Delta T\). A sequential training strategy is employed. At each step, a PINN model is trained on the domain \(\Omega \times I_i\). The initial condition for this domain at time \(t_i\) is obtained from the predictions of the previously trained PINN model. For the first time window, the exact initial condition is used. However, for subsequent time windows, the initial conditions are derived from the predictions of the preceding PINN model.  

The S2S approach has proven to be an effective strategy for handling stiff PDEs, such as fourth-order models \cite{mattey_novel_2022} and transport phenomena with high-speed values \cite{penwarden_unified_2023}. Inspired by these studies, we adopt the S2S approach to train our PINN model for level set problems, aiming to enhance the accuracy.

\subsubsection{SOAP optimizer}
\label{soap}
Recently, \cite{wang_gradient_2025} have shown that traditional first order optimizer like Adam struggle with the composite loss functions of PINNs. The simultaneous optimization of multiple loss terms makes it difficult to align the gradients in the same direction, resulting in competing loss terms and poor optimization. Shampoo with Adam in the Preconditioner’s eigenbasis or SOAP \cite{wang_gradient_2025,vyas_soap_2025} is a second-order quasi-Newton optimizer that naturally aligns gradients to provide better and faster optimization.
SOAP is especially effective when solving coupled problems with multiple loss terms.

\subsection{PINNs for the level set equation}
\label{pinns_levelset}

In our proposed PINN solver for level set based problems, the loss function contains the initial and residual losses:
\begin{equation}
   \mathcal{L}_{level\, set}=\lambda_{ic}\mathcal{L}_{ic}+\lambda_{r}\mathcal{L}_{r}.
    \label{eq:loss_levelset}
\end{equation}
The level set residual is defined as:
\begin{equation}
   \mathcal{L}_{r}(\bm{x}, t) = \frac{\partial \phi}{\partial t}(\bm{x}, t) + \bm{v}(\bm{x},t) \cdot \nabla \phi(\bm{x}, t),
   \label{eq:ls_res}
\end{equation}
where \(\bm{v}(\bm{x},t)\) is the imposed velocity field.
We note that the boundary conditions are not enforced in the level set benchmark problems.

In the following, we propose two additional loss terms that can be incorporated into the level set PINN loss function. These include an \textit{Eikonal loss}, designed to enforce the Eikonal property \eqref{eq:ls_grad1}, and a \textit{mass loss}, aimed at improving mass conservation of the PINN model.

\subsubsection{Eikonal loss}
\label{eik_loss}
For a given level set function \(\phi\), the Eikonal residual is defined as:  
\begin{equation}
   \epsilon_{eik}(\bm{x}, t) = \| \nabla \phi(\bm{x}, t) \|^2 - 1,
   \label{eq:ls_eik_res}
\end{equation}
where the residual represents the deviation from the Eikonal property. The Eikonal loss is then computed at selected residual points $(\bm{x}_{eik}^i, t_{eik}^i)$ as follows:  
\begin{equation}
    \mathcal{L}_{eik}(\theta) = \frac{1}{N_{eik}} \sum_{i=1}^{N_{eik}} |\epsilon_{eik}(\bm{x}_{eik}^i, t_{eik}^i)|^2,
    \label{eq:loss_eik_term}
\end{equation}
where \(N_{eik}\) is the number of residual points used to evaluate the Eikonal loss. 
The updated loss function can be rewritten as:
\begin{equation}
   \mathcal{L}(\theta)=\lambda_{ic}\mathcal{L}_{ic}(\theta) + \lambda_{r}\mathcal{L}_r(\theta) + \lambda_{eik}\mathcal{L}_{eik}(\theta).
   \label{eq:ls_loss_eik}
\end{equation}
By incorporating this term directly into the training process of the PINN, our approach eliminates the need for a separate reinitialization step to enforce the signed distance property. This integration not only simplifies the workflow but also offers a significant advantage over traditional numerical techniques, where such initialization is typically required.

\subsubsection{Mass loss}
\label{mass_loss}
The second loss term that can be added is the mass loss, which involves calculating the mass of phase \(i\) at each time step and constraining the PINN model to maintain it equal to the initial mass. In our study of two-phase problems, we implement the mass loss by calculating the area of \(\Omega_1\) and enforcing it to remain constant over time. The area is computed using the Monte Carlo (MC) sampling approach.
In this context, the area \(\mathcal{A}_\theta\) is estimated by sampling \(N\) random points \(\bm{x}_i\) uniformly from the whole domain \(\Omega\) and checking whether they belong to the region \(\Omega_1\), defined as the subset of \(\Omega\) where the level set function \(\phi(\bm{x}, t) < 0\):  
\begin{equation}
\Omega_1(t) = \{ \bm{x} \in \Omega \ | \ \phi(\bm{x}, t) < 0 \}.
\end{equation}
The area \(\mathcal{A}_\theta\) is then computed as:
\begin{equation}
  \mathcal{A}_\theta(t) \approx \frac{\text{Number of points in } \Omega_1(t)}{N} \cdot \mathcal{V}(\Omega),
\end{equation}
where \(\mathcal{V}(\Omega)\) is the volume of the sampling domain \(\Omega\).

Given that the exact area at \(t = 0\), denoted as \(\mathcal{A}_{exact}\), must be conserved over time, the mass loss is calculated as:
\begin{equation}
    \mathcal{L}_{mass}(t_i, \theta) = (\mathcal{A}_{\theta}(t_i) - \mathcal{A}_{exact})^2,
    \label{eq:loss_mass_term}
\end{equation}
where \(\mathcal{A}_{\theta}(t_i)\) is the predicted area at specific time points \(t_i \in [0, T]\). To compute the overall mass loss, the mean of \(\mathcal{L}_{mass}(t_i, \theta)\) across all sampled time points is used:
\begin{equation}
    \mathcal{L}_{mass}(\theta) = \frac{1}{N_t} \sum_{i=1}^{N_t} \mathcal{L}_{mass}(t_i, \theta),
    \label{eq:mean_mass_loss}
\end{equation}
where \(N_t\) is the total number of sampled time points. This approach ensures that the model consistently conserves the area for all selected time intervals.

Two methods were tested to implement the mass loss in the loss function.
\textit{Method 1}: is straightforward and involves directly adding the mass loss to the composite loss function:
\begin{equation}
   \mathcal{L}(\theta)=\lambda_{ic}\mathcal{L}_{ic}(\theta) + \lambda_{r}\mathcal{L}_r(\theta) + \lambda_{mass}\mathcal{L}_{mass}(\theta).
   \label{eq:ls_mass_loss_1}
\end{equation}
While simple, this method can be computationally expensive as the mass loss needs to be computed at every training step. This involves computing the level set field solution for the complete spatio-temporal domain which becomes costly when performed several thousand times. Therefore, a more efficient approach is desirable to limit the frequency of mass loss computation.

\textit{Method 2}: addresses this limitation by using a two-network framework, which involves computing the mass loss in a second distinct network initialized with transfer learning.
The first network is trained without the mass loss term. The trained parameters of the first network are then used to initialize a second network. This second network is trained exclusively on the mass loss term. The model is then evaluated with the saved parameters of the second network. Both loss functions are defined as:
\begin{equation}
    \begin{aligned}
           \mathcal{L}_{net\, 1} &=\lambda_{ic}\mathcal{L}_{ic}+\lambda_{r}\mathcal{L}_{r},\\
           \mathcal{L}_{net\, 2} &=\mathcal{L}_{mass}.
    \end{aligned}
    \label{eq:ls_mass_loss_2}
\end{equation}
By decoupling the training, \textit{Method 2} reduces the computational burden and allows for a tailored number of training steps for mass loss. This method also has the advantage of allowing us to choose a custom amount of collocation points.

\subsection{Algorithm implementation}

The code was built on JAX \cite{bradbury_jax_2018} along with the JAXPI \cite{wang_experts_2023} library, which allows for easy implementation of PINNs, \textit{PirateNets}, and the improvements mentioned in Section \ref{sec:Improved_PINNs}. The runs were tracked with Weights \& Biases  (wandb) \cite{biewald_weights_2020} and the visualizations were done with Matplotlib \cite{hunter_matplotlib_2007}. The results were computed using a single NVIDIA V100 GPU on one of the Digital Research Alliance of Canada's compute clusters. The code will be made available upon request.

\section{Results}
\label{results}

In the following sections, we present the numerical results obtained from the proposed PINN solver for two well-known level set benchmarks. Before diving into these benchmarks, we validate our PINN framework using two challenging benchmark problems in fluid dynamics, the 1D and 2D Burgers' equations \cite{burgers_mathematical_1948}. To facilitate reading, this section, including ablation studies, was placed in \ref{app:burgers}. Next, we focus on two level set benchmarks: Zalesak's disk and the time-reversed vortex flow. We finish with a complex application: coupling the level set equation with the incompressible Navier-Stokes equations for modelling of a 2D dam break problem in two-phase flow.

To evaluate the accuracy of the PINN solver, we compare its performance against a reference solution using the relative \( L^2 \) error metric:

\begin{equation}
 \text{Relative } L^2 \text{ Error} = \frac{\| u_{\text{PINN}} - u_{\text{reference}} \|_2}{\| u_{\text{reference}} \|_2},
 \label{eq:l2error}
\end{equation}
where \( u_{\text{PINN}} \) is the solution obtained from the PINN solver, and \( u_{\text{reference}} \) is the high-fidelity reference solution.

In this section, we define the following configurations:
\begin{itemize}
    \item \textit{Plain}: the original PINN formulation \cite{raissi_physics-informed_2019},
    \item \textit{Default}: the improved PINN approach \cite{wang_experts_2023},
    \item \textit{\textit{PirateNet}}: PirateNet architecture including the improvements from \textit{Default}.\cite{wang_piratenets_2025},
    \item \textit{Sota}: PirateNet with the optimal hyperparameters, determined through a detailed hyperparameter sweep using WandB's Bayesian sweep approach \cite{biewald_weights_2020}.
\end{itemize}

\subsection{Level set tests}
\label{levelset}
In the subsequent sections, we present a series of level set benchmark tests to rigorously validate the effectiveness and reliability of the proposed PINN approach.

\subsubsection{Illustrative example}
\label{cercle}
In the following, we construct a quasi-level set problem to evaluate the efficiency of the vanilla PINN approach and investigate its behavior and accuracy during long-time integration within the level set framework. 

We consider the transport of a circle with a radius of \( r = 0.15 \) within a square domain, \([0, 1] \times [0, 1]\). The circle is initially centered at \( (c_x, c_y) = (0.17, 0.5) \) and is transported with a horizontal velocity of \( v_x = 1.0 \). As the circle moves along the horizontal direction, its velocity is reversed upon reaching the domain boundaries, ensuring it remains confined within the square domain and exhibits oscillatory motion. 

The circle represents an interface separating two fluids, and it is required to preserve its initial shape without deformation throughout the simulation. The exact position of the circle at any given time can be determined using the analytical solution of the corresponding transport equation.

Next, we adopt the PINN approach to solve this transport problem. We utilize an MLP with 4 hidden layers, each containing 256 neurons. Due to the discontinuous nature of the velocity parameter \( v_x \), employing a single time window did not produce satisfactory results. Consequently, we adopt a sequence-to-sequence approach, where each window has a constant velocity. The PINN model is trained for \( 2 \times 10^4 \) steps in each window using the Adam optimizer with an initial learning rate of \( 1 \times 10^{-3} \). The learning rate is decayed exponentially at every \( 1 \times 10^3 \) steps. It is important to note that no enhancements to PINN training, as discussed in Section \ref{sec:Improved_PINNs}, were applied in this example.

\begin{figure}[!htb]
 \begin{minipage}[t]{0.78\textwidth}  
  \centering
  \textbf{Impact of the Eikonal regularization}\par\medskip
  \includegraphics[width=\textwidth]{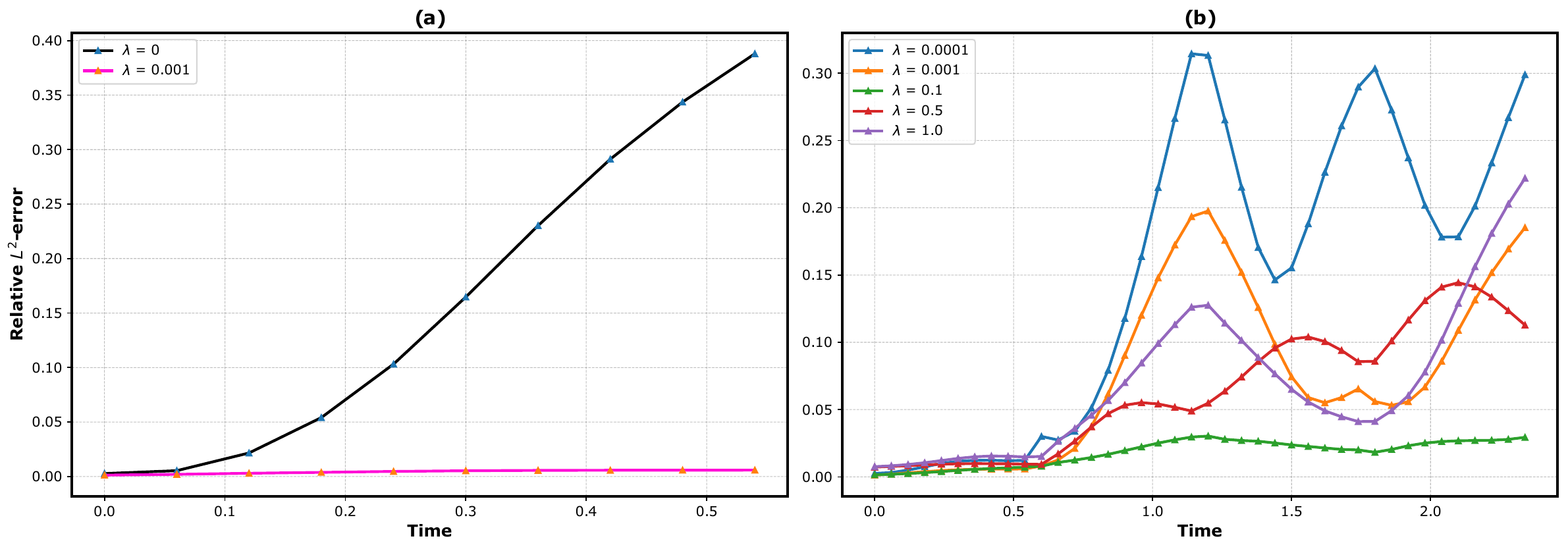}
 \end{minipage}
  \hspace{-0.3cm}
   \begin{minipage}[t]{0.8\textwidth}  
   \centering
   \textbf{Impact of the Mass regularization}\par\medskip
   \includegraphics[width=\textwidth]{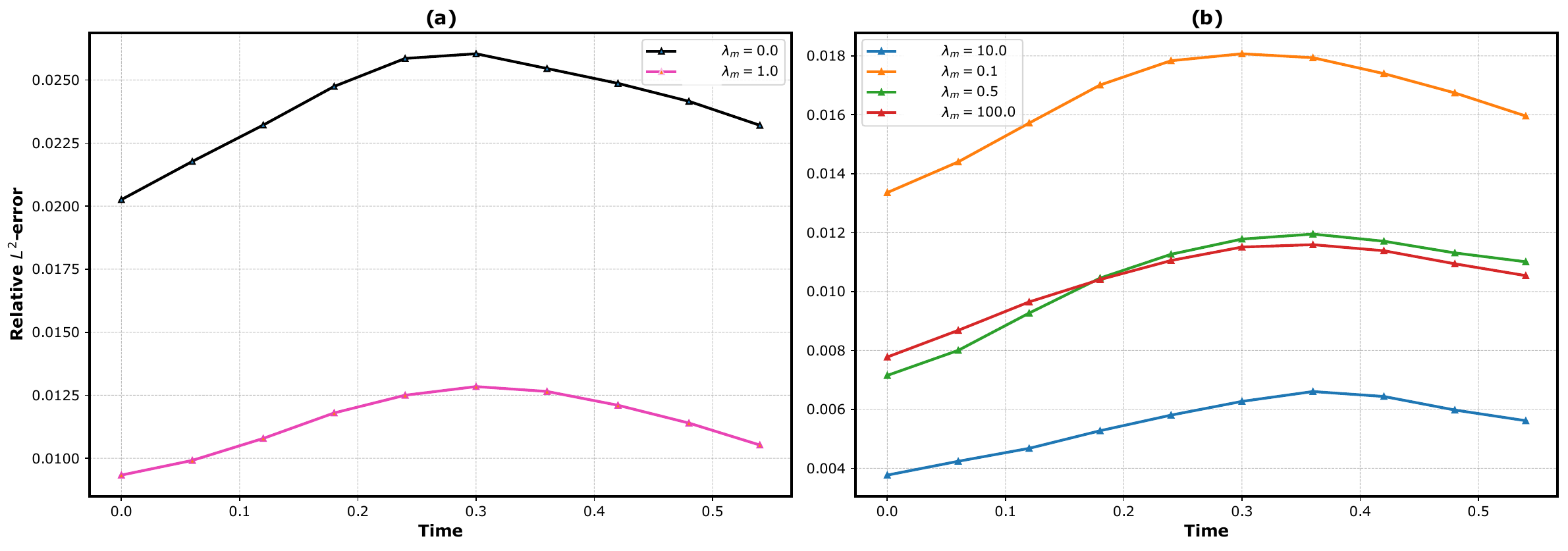}
 \end{minipage}
   \hspace{-0.3cm}
   \begin{minipage}[t]{0.9\textwidth}  
   \centering
   \textbf{Obtained solution}\par\medskip
   \includegraphics[width=\textwidth]{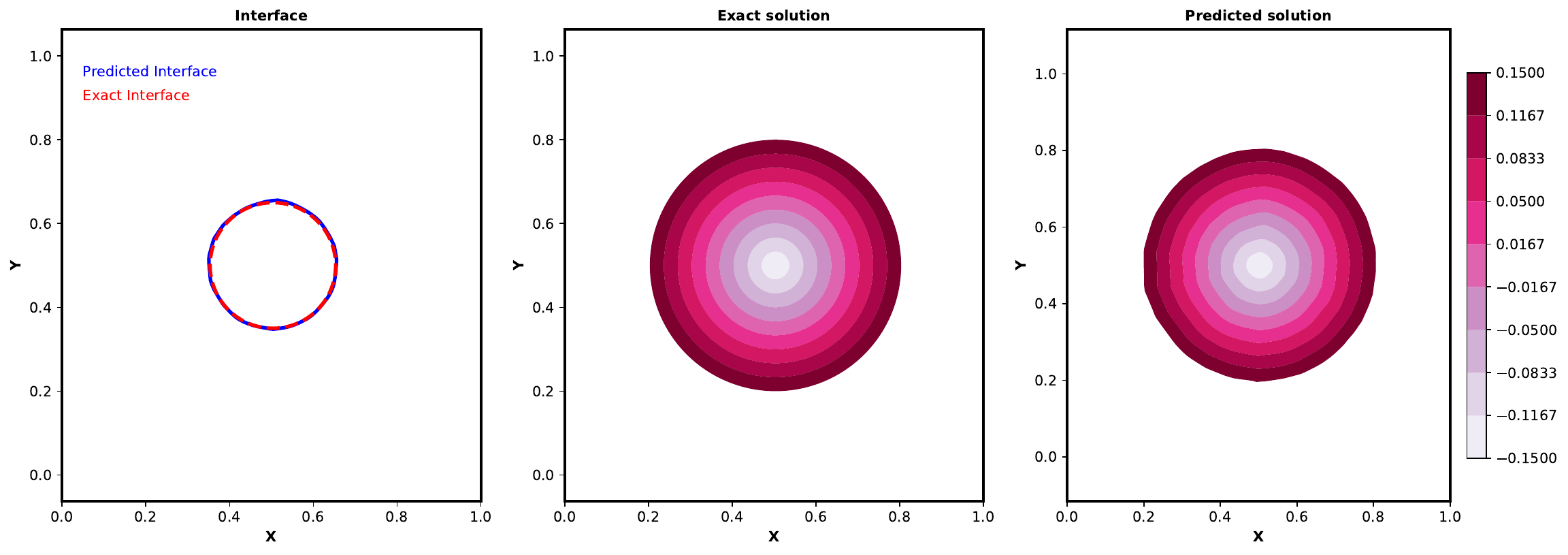}
 \end{minipage}
\caption{Visualization of the PINN results: (Top) Profiles showing the impact of Eikonal  regularization. (Middle) The impact of  mass regularization. (Bottom) The obtained solution for the transport of the circular interface in the square domain.}
\label{results_circle}
\end{figure}

The obtained results are illustrated in Figure \ref{results_circle}. First, without applying Eikonal or mass regularization, it is evident that the error increases as the simulation time progresses. During the simulation using the vanilla PINN approach, the circular interface completely disappears after just two periods of oscillation due to numerical error accumulation.  

Incorporating the Eikonal loss with a relatively small weight, on the order of \(10^{-1}\), significantly improves the results, reducing the error over time by approximately four orders of magnitude. A similar improvement is observed when the mass regularization term is applied. However, the most critical aspect is the judicious selection of the weights corresponding to these regularization terms. For instance, assigning an excessively large weight to the Eikonal loss may lead the PINN solver to prioritize satisfying the Eikonal equation rather than the transport equation, thereby deviating from the intended solution.  

From this simple example, we conclude that when using a vanilla PINN approach for long-time integration of interface motion within the level set framework, it is essential to incorporate both mass conservation and the Eikonal property to achieve efficient and accurate simulations of the problem.

\subsubsection{Level set Zalesak's disk}
\label{levelset_zalesak}
Next, we explore a popular benchmark problem for testing level set methods, the rigid body rotation of Zalesak's disk \cite{zalesak_fully_1979,rider_reconstructing_1998}. This test involves submitting a non-deforming interface to the following rotating flow field:
\begin{equation}
    u(x,y)=\frac{2\pi}{T}(0.5-y)\quad \text{and} \quad v(x,y)=\frac{2\pi}{T}(x-0.5),
    \label{eq:zalesak_flow_field}
\end{equation}
with \(T=2\pi\) s. The computational domain is a circle of radius \(R = 0.5\), centered at (0.5, 0.5) within a unit square. The interface consists of a slotted disk centered at \((0.5,0.75)\) with radius \(r = 0.15\). The slot is aligned along the disk's centerline and extends vertically with a height \(h = 0.25\) and width \(w = 0.05\).

The circular domain is chosen to avoid numerical issues related to the inflow boundary condition when solving the reference solution with a finite element method (FEM). The reference solution was obtained using the SUPG-stabilized finite element method, with a quadratic approximation for the level set function on unstructured triangular elements (T6) and a third-order explicit strong–stability–preserving Runge–Kutta (SSP–RK) time integration scheme \cite{fahsi_numerical_2017} implemented with MATLAB. Furthermore, the mesh used for the FEM solver had \(10470\) elements, which enabled us to interpolate the results onto a \(101\times101\) uniformly spaced Python-compatible grid. The FEM simulation lasted \(t_{end}=T\), and the time step size was \(\Delta t_{FEM} = 1.57 \times  10^{-3}\) s. The temporal resolution was downsampled to  \(\Delta t_{PINN} = 1.57\times 10^{-1}\) s for comparison of FEM and PINN models.

We trained a set of PINN configurations (\textit{Plain}, \textit{Default}, \textit{PirateNet}) to learn the evolution of the interface. The configurations follow the same standard as in previous sections. Their detailed hyperparameters can be found in Table \ref{tab:ls_zalesak_hp}.

\begin{figure}
    \centering
    \includegraphics[width=0.6\linewidth]{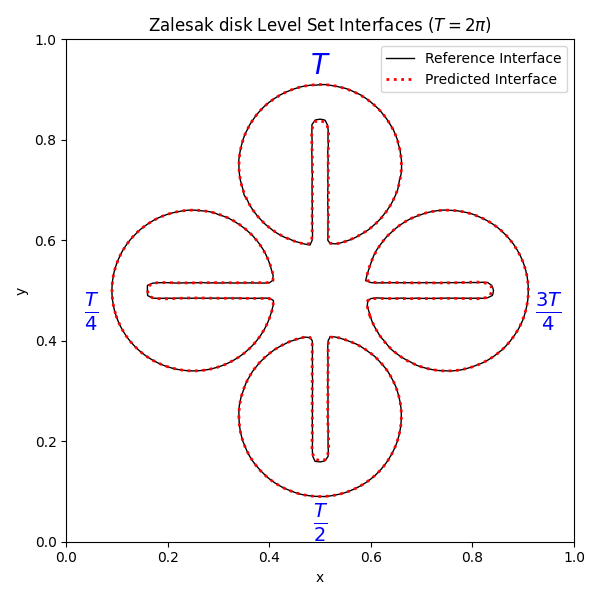}
    \caption{Zalesak's disk \textit{reference} and \textit{Sota} solution evolution (\(T=2\pi \text{ s}\))}
    \label{fig:zalesak_ts}
\end{figure}

From Table \ref{tab:ls_zalesak_l2_basic}, we can again see that \textit{PirateNet} performs better than \textit{Default} and \textit{Plain} with a relative error \(L^2_{PirateNet}=0.35\%\) compared to \(L^2_{Default}=2.96\%\) and \(L^2_{Plain}=4.18\%\). With the PirateNet architecture, we then performed a Bayesian hyperparameter sweep using Wandb's sweep module \cite{biewald_weights_2020}. Table \ref{tab:ls_zalesak_hp}'s \textit{Sota} configuration summarizes the optimal hyperparameters, allowing us to lower the error down to \(L^2_{Sota}=0.14\%\). 
Figure \ref{fig:zalesak_ts} for the \textit{Sota} configuration shows the graphical results at each simulation quarter, where we can confirm that the interface is well captured.
\begin{table}[!h]
    \centering
        \begin{tabular}{c|c}
            \hline
             Configuration name& \(L^{2}\) error (\%)\\
             \hline
             Plain& 4.18\\
             Default& 2.96\\
             PirateNet& 0.35\\
             Sota& \textbf{0.14}\\
        \end{tabular}
    \caption{Level set Zalesak's disk: \(L^2\) error of the output \(\phi\) field}
    \label{tab:ls_zalesak_l2_basic}
\end{table}

Before moving on, we validate our definition from section \ref{piratenet_arch} where we define the number of layers in a \textit{PirateNet} as being equal to its amount of residual blocks. Given that each \textit{PirateNet} residual block contains three dense layers, we want to make sure that we get better results with \textit{PirateNet} not only because the model has more parameters. We do so by testing the influence of model depth and size on the error. We tested \textit{Plain} and \textit{Default} with nine layers and \textit{PirateNet} with one layer (one residual block). We can find these results in Table \ref{tab:ls_zalesak_size_influence} along with the benchmark three layers of \textit{PirateNet}.

From Table \ref{tab:ls_zalesak_size_influence}, we notice that \textit{Plain} and \textit{Default} improve with deeper networks, but \textit{PirateNet} still offers the best result. Indeed, \textit{PirateNet} with one layer has a slight increase in error compared to the three-layer version but remains the best configuration while being less than half the size of the nine-layer \textit{Plain} and \textit{Default}. 

Given these results, we will keep our definition of a PirateNet layer as equal to a PirateNet residual block for further experiments.
\begin{table}[!h]
    \centering
    \resizebox{\textwidth}{!}{
        \begin{tabular}{c|cccc}
            \hline
            Configuration name & 
            \shortstack{Amount \\ of layers} & 
            \shortstack{Amount of \\ parameters ($10^3$)} & 
            \shortstack{Model \\ size (MB)} & 
            \shortstack{\(L^{2}\) error \\ (\%)} \\
            \hline
            Plain & 9 & 527 & 2.01 & 2.56 \\
            Default & 9 & 727 & 2.77 & 0.59 \\
            PirateNet & 1 & 330 & 1.26 & 0.43 \\
            PirateNet & 3 & 727 & 2.77 & 0.35 \\
            \hline
        \end{tabular}
    }
    \caption{Level set Zalesak's disk: Model depth and size influence on the error.}
    \label{tab:ls_zalesak_size_influence}
\end{table}

\begin{figure}[!h]
    \centering
    \includegraphics[width=0.6\linewidth]{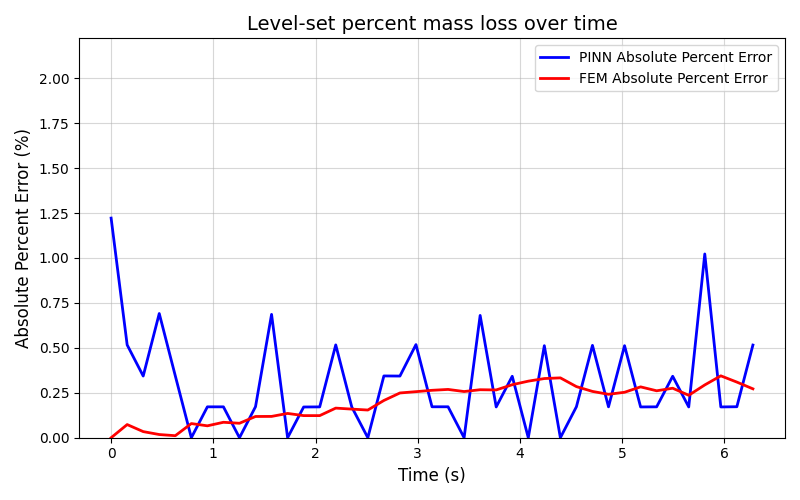}
    \caption{Zalesak's disk absolute percent mass error evolution (\(T = 2\pi \text{ s}\))}
    \label{fig:zalesak_mape}
\end{figure}

Using the trained weights of \textit{Sota}, we quantified the mass loss throughout the simulation. The reference mass is calculated trivially using the disk's area minus the slot's area at \(t=0\),  which should remain constant throughout the simulation. We compared the evolution of mass loss of \textit{Sota} and the \textit{FEM} solver in Figure \ref{fig:zalesak_mape}. We notice that the PINN solver has a higher variance, while \textit{FEM} follows a more constant error accumulation pattern. This is expected since PINNs solve the whole spatial-temporal grid at once, while \textit{FEM} uses a time-marching scheme. Nonetheless, \textit{PirateNet} demonstrates here that it can effectively capture mass patterns with similar results to a high-precision level set FEM solver. Their respective mean absolute percent mass error are \(MAPE_{Sota }= 0.31\%\) and \(MAPE_{FEM} = 0.21\%\).

After assessing the performances of a basic \textit{PirateNet} on Zalesak's disk, we decided to test how the proposed enhancements from Section \ref{pinns_levelset} would affect the results.

For the Eikonal loss term, we tested several weighting methods. We started by including the weight \(\lambda_{eik}\) inside the gradient normalization scheme using the same initial weight as the other terms: \(\lambda_{eik} = \lambda_{ic} = \lambda_{r} = 1\). Even with a frequent weight update every 500 training steps, this worsened the solution by almost two orders of magnitude compared to the benchmark \textit{PirateNet}. After trying a lower initial weight of \(\lambda_{eik}=0.01\), the results were better but still one order of magnitude worse than the benchmark. We also noticed that the gradient normalization tended to increase the Eikonal loss weight throughout the training to values that were higher than the other weights, further worsening the solution. 

\begin{figure}
    \centering
    % First subfigure
    \begin{subfigure}[b]{0.32\linewidth}
        \centering
        \includegraphics[width=\linewidth]{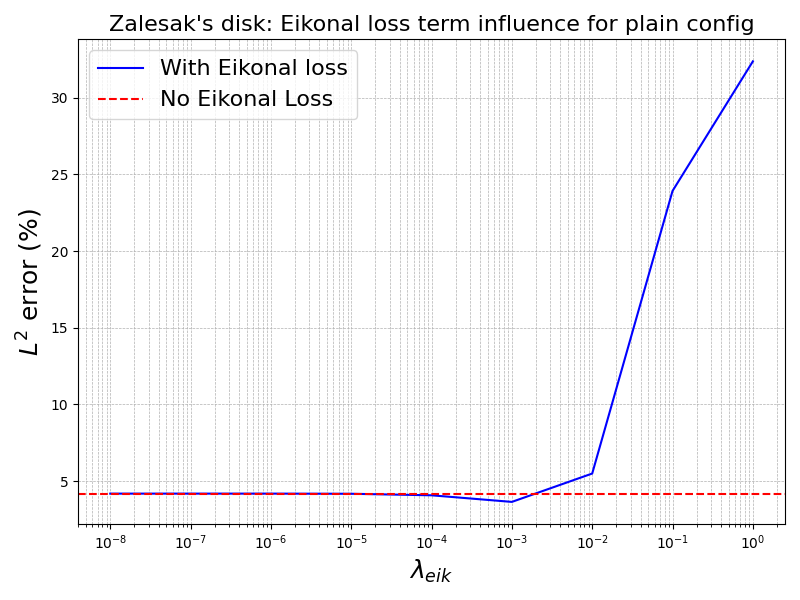}
        \caption{Influence of \(\lambda_{eik}\) on \textit{Plain}}
        \label{fig:zalesak_eik_plain}
    \end{subfigure}
    %\hfill
    % Second subfigure
    \begin{subfigure}[b]{0.32\linewidth}
        \centering
        \includegraphics[width=\linewidth]{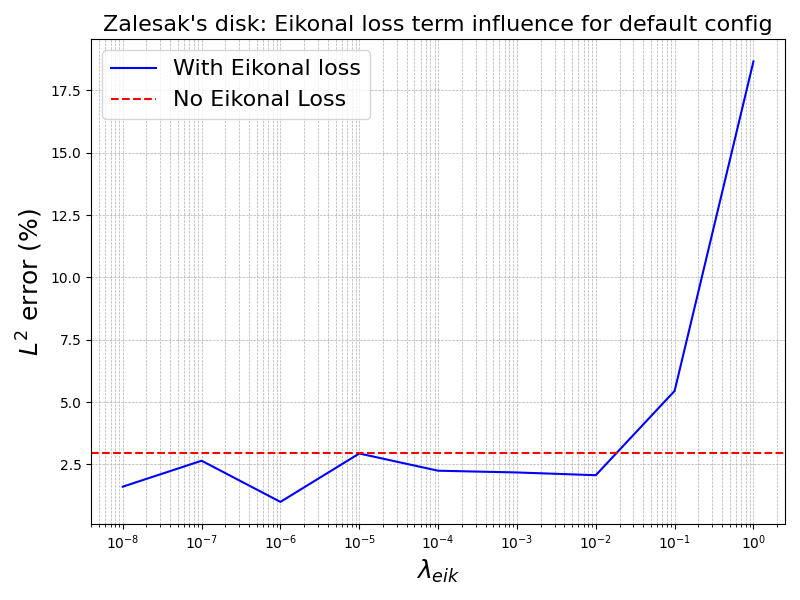}
        \caption{Influence of \(\lambda_{eik}\) on \textit{Default}}
        \label{fig:zalesak_eik_default}
    \end{subfigure}
    %\hfill
    % Third subfigure
    \begin{subfigure}[b]{0.32\linewidth}
        \centering
        \includegraphics[width=\linewidth]{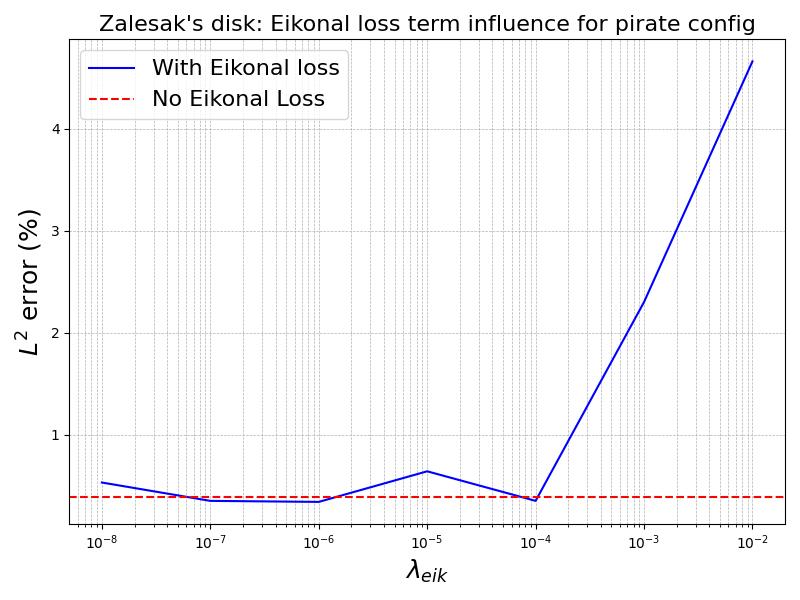}
        \caption{Influence of \(\lambda_{eik}\) on \textit{PirateNet}}
        \label{fig:zalesak_eik_pirate}
    \end{subfigure}
    \caption{Zalesak's disk: influence of the Eikonal loss term \(\lambda_{eik}\).}
    \label{fig:zalesak_eik_influence}
\end{figure}

Given this, we consider the Eikonal loss as a regularization term that is kept constant throughout the training and excluded from the gradient normalization-based weight updates. We tested a range of weights, \(\lambda_{eik}\), and measured their influence on the error. From Figure \ref{fig:zalesak_eik_influence}, we can see that adding the Eikonal loss offers no significant improvement to the solution of Zalesak's disk rigid body rotation problem. We also notice that higher weights above \(10^{-3}\) negatively impact the solution. However, as seen with the \textit{Default} config, we note that under certain circumstances the Eikonal loss can slightly improve the results with a carefully chosen weight \(\lambda_{eik}\).

Next, we tested the influence of adding a mass loss employing \textit{Method 2} from Section \ref{mass_loss}. \textit{Method 1} was ignored for this case because of its computational overhead. \textit{Method 2}'s approach entails training a network exclusively with a mass loss, using the fully trained weights of a network initially trained with the standard initial condition and residual losses. 
Using \textit{Sota}'s weights, we trained the network for a further 1000 iterations using a batch size of \(64 \times 64\) spatial points over all time steps to compute the output field's prediction. The other hyperparameters were the same as \textit{Sota}'s. Adding this mass loss did not improve the error for Zalesak's disk problem, as the error increased from \(L^2_{Sota} = 0.14\%\) to \(L^2_{Sota+mass} = 0.29\%\). Furthermore, the mean absolute percent mass error increased from \(MAPE_{Sota}=0.31\%\) to \(MAPE_{Sota+mass} = 1.09\%\).
As for regularization methods, adding the Eikonal and mass loss terms with well-chosen weights may slightly affect the accuracy but provide a stabilization enhancement.

\subsection{Level set vortex flow}
\label{levelset_vortex}
This section will show the capabilities of \textit{PirateNet} to solve a benchmark level set problem as defined in \cite{rider_reconstructing_1998}, the time-reversed vortex flow. This test involves stretching and spinning a circle within the first half-period. During the second half, the flow is reversed and the circle returns to its initial state. This will test the \textit{PirateNet}'s ability to learn the level set function \(\phi\) under a varying velocity field with strong deformation.
The definition of a single vortex flow's stream function is 
\begin{equation}
    \Psi_{single \, vortex}(x,y) = \frac{1}{\pi}\sin^2(\pi x)\sin^2(\pi y)
    \label{eq:single_vortex}
\end{equation}
From Eq. \ref{eq:single_vortex}, we can get the time-reversed vortex flow stream function by multiplying by \(\cos \left(\dfrac{\pi t}{T}\right)\), where \(T\) is the period.
\begin{equation}
    \Psi_{time-reversed \, vortex}(x,y, t)=\frac{1}{\pi}\sin^2(\pi x)\sin^2(\pi y)\cos \left(\dfrac{\pi t}{T}\right)
    \label{eq:time_rev_vortex}
\end{equation}
From Eq. \ref{eq:time_rev_vortex}, we can derive the flow fields with:
\begin{equation}
    u=-\frac{\partial\Psi}{\partial y} \quad \text{and} \quad v=\frac{\partial\Psi}{\partial x}
    \label{eq:uv_fields}
\end{equation}
where
\begin{equation}
    u_{time-reversed \, vortex}(x,y,t)=-2 \sin^2(\pi x)\cos(\pi y)\sin(\pi y)\cos\left(\dfrac{\pi t}{T}\right),
    \label{eq:u_vortex}
\end{equation}
and
\begin{equation}
    v_{time-reversed \, vortex}(x,y,t)=2\cos(\pi x)\sin(\pi x)\sin^2(\pi y)\cos\left(\dfrac{\pi t}{T}\right).
    \label{eq:v_vortex}
\end{equation}
We used the same period \(T = 8\) s as in \cite{toure_stabilized_2016} for our tests. The computational domain was defined as \(x,y \in [0,1]\), and the computational domain was split into two subdomains with a circle or radius of \(r = 0.15\), centered at \((0.5,0.75)\).
For this problem, no boundary conditions are enforced. This implies that the boundary condition loss term \(\mathcal{L}_{bc}\) is removed from Eq. \ref{eq:loss_function}.

The goal is to learn the output field of the level set function \(\phi\).
The reference solution was generated using the same FEM solver as in Section \ref{levelset_zalesak}.

We first tested our three main model configurations, the \textit{Plain}, the \textit{Default}, and the \textit{PirateNet} on the time-reversed vortex flow problem. These configurations represent the use of the same features as in Sections \ref{burgers1d} to \ref{levelset_zalesak}. They were compared using the same basic hyperparameters as seen in Table \ref{tab:ls_vortex_hp}. A particularity with the time-reversed vortex flow is that the flow reverses at the half-period. Therefore, we need to use an S2S scheme with two time windows to ensure the gradients are reset with the proper signs when the flow reverses. 

\begin{figure}
    \centering
    \includegraphics[width=0.7\linewidth]{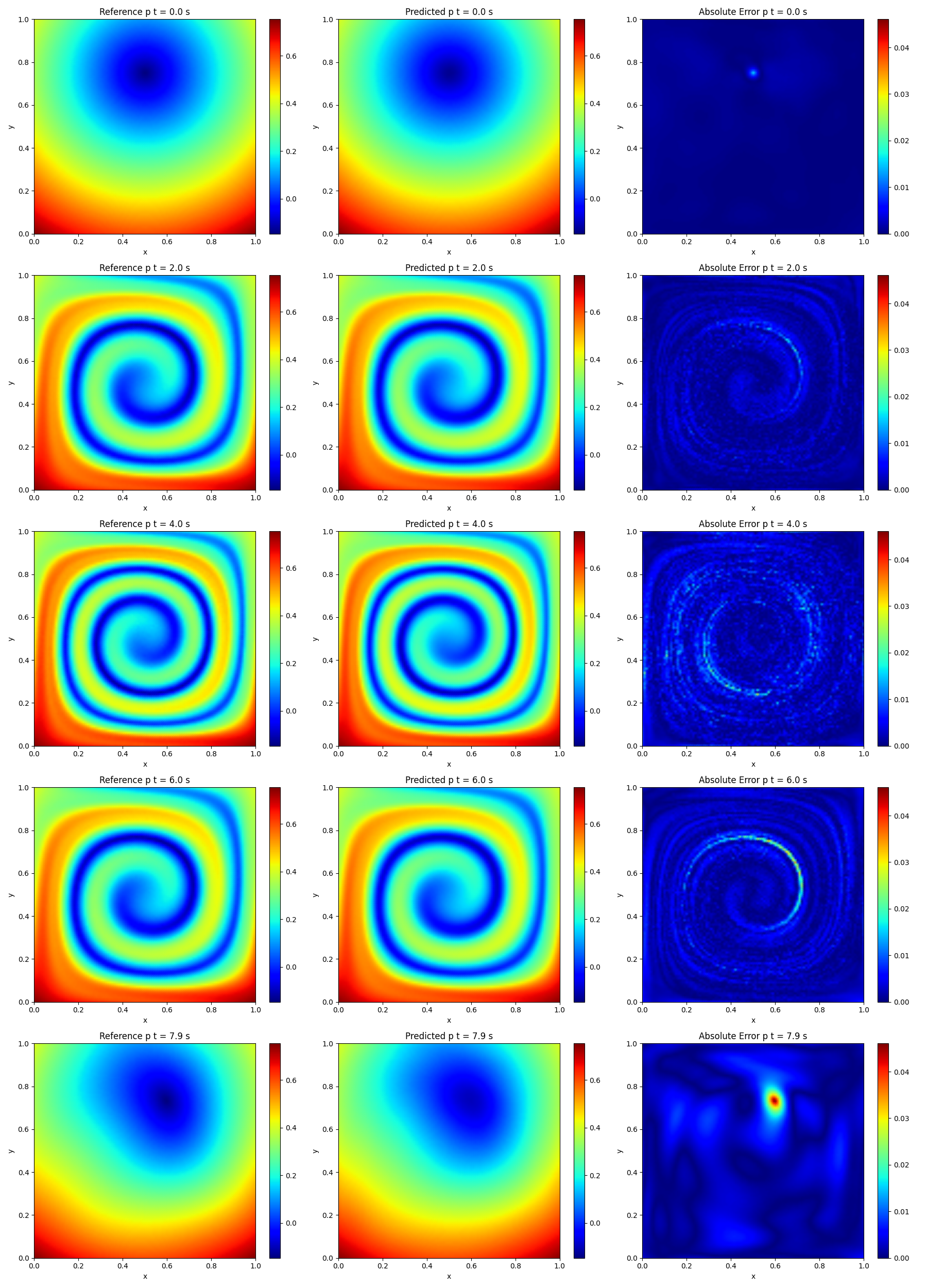}
    \caption{Level set vortex: Reference, predicted, and absolute error of \(\phi\) fields at different time steps.}
    \label{fig:levelset_phi}
\end{figure}

From Table \ref{tab:ls_vortex_l2_basic}, we can see that the \textit{PirateNet} is the one that performed the best with \(L^{2}_{PirateNet} = 5.24\%\). On the other hand, the \textit{Plain} and \textit{Default} configurations lacked expressive capacity, as shown by their respective errors \(L^{2}_{Plain} = 51.20\%\) and  \(L^{2}_{Default} = 20.86\%\).
With the \textit{PirateNet} architecture, we then performed a Bayesian hyperparameter sweep using Wandb's sweep module \cite{biewald_weights_2020}.  Table \ref{tab:ls_vortex_hp}'s \textit{Sota} configuration shows the parameters that obtained the best results from this sweep. With \textit{Sota}, we were able to reduce the error down to \(L^2_{Sota}=0.85\%\). 
\begin{table}
    \centering
    %\resizebox{0.4\textwidth}{!}{
        \begin{tabular}{c|c}
            \hline
             Configuration name& \(L^{2}\) error (\%)\\
             \hline
             Plain& 51.20\\
             Default& 20.86\\
             PirateNet& 5.24\\
             Sota&\textbf{0.85}\\
        \end{tabular}
    %}
    \caption{Level set vortex: \(L^2\) error of the output \(\phi\) field.}
    \label{tab:ls_vortex_l2_basic}
\end{table}
The graphical results for \textit{Sota} at five different time steps are shown in Figure \ref{fig:levelset_phi}.

\begin{figure}
    \centering
    \includegraphics[width=0.7\linewidth]{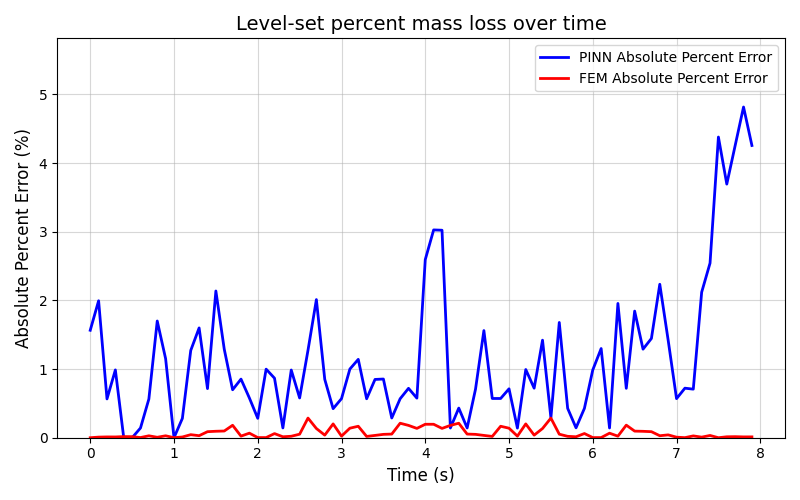}
    \caption{Level set vortex: Evolution of mass loss with fully trained \textit{Sota}.}
    \label{fig:levelset_sota_mass}
\end{figure}

We then quantified the loss of mass during the simulation using the area of the circle of radius \(r = 0.15\) as a reference mass. The evolution of the absolute percent mass loss for \textit{Sota} and the reference \textit{FEM} is shown in Figure \ref{fig:levelset_sota_mass}. The mass loss is higher for \textit{Sota} than for \textit{FEM}, but it remains reasonable. 
 Their respective mean absolute percent mass errors are \(MAPE_{Sota} = 1.18\%\) and \(MAPE_{FEM} = 0.07\%\).

As done with Zalesak's disk in Section \ref{levelset_zalesak}, we tested the effect of the proposed enhancements from section \ref{pinns_levelset}, but this time on the more challenging time-reversed vortex flow problem.

\begin{figure}
    \centering
    % First subfigure
    \begin{subfigure}[b]{0.32\linewidth}
        \centering
        \includegraphics[width=\linewidth]{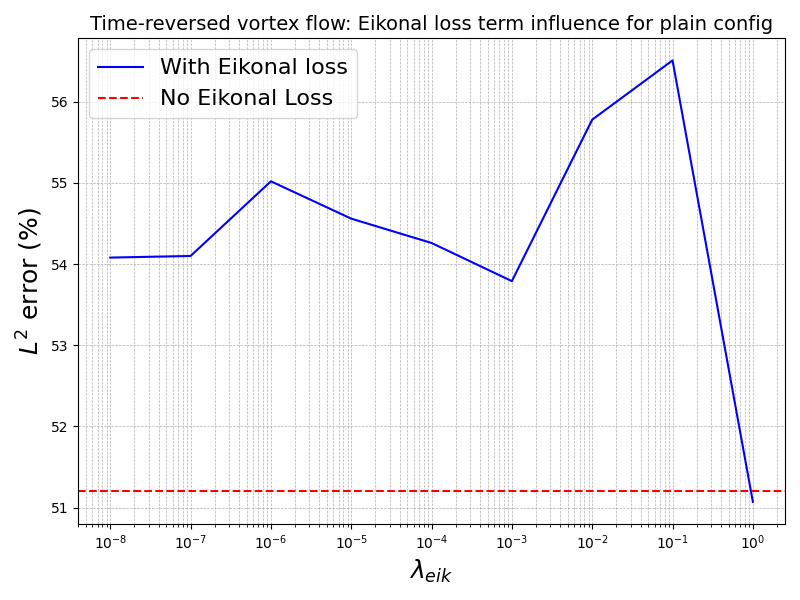}
        \caption{Influence of \(\lambda_{eik}\) on \textit{Plain}}
        \label{fig:vortex_eik_plain}
    \end{subfigure}
    %\hfill
    % Second subfigure
    \begin{subfigure}[b]{0.32\linewidth}
        \centering
        \includegraphics[width=\linewidth]{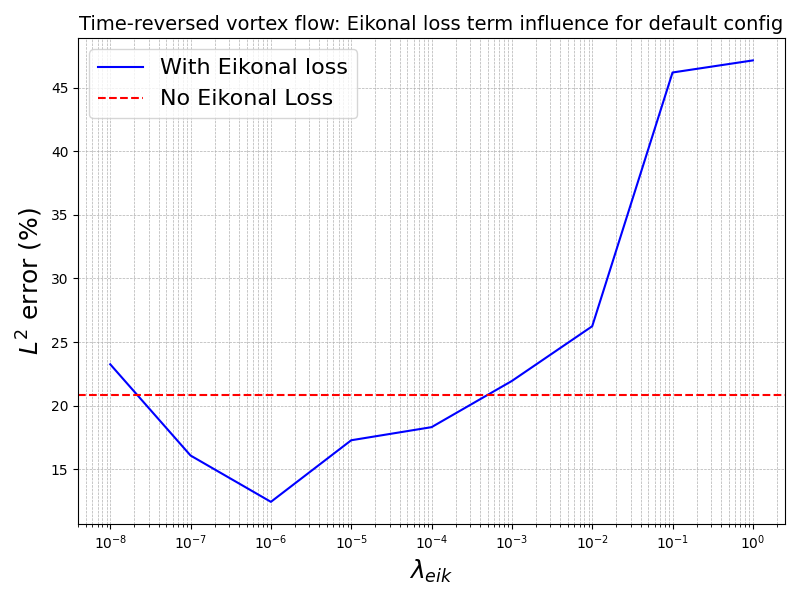}
        \caption{Influence of \(\lambda_{eik}\) on \textit{Default}}
        \label{fig:vortex_eik_default}
    \end{subfigure}
    %\hfill
    % Third subfigure
    \begin{subfigure}[b]{0.32\linewidth}
        \centering
        \includegraphics[width=\linewidth]{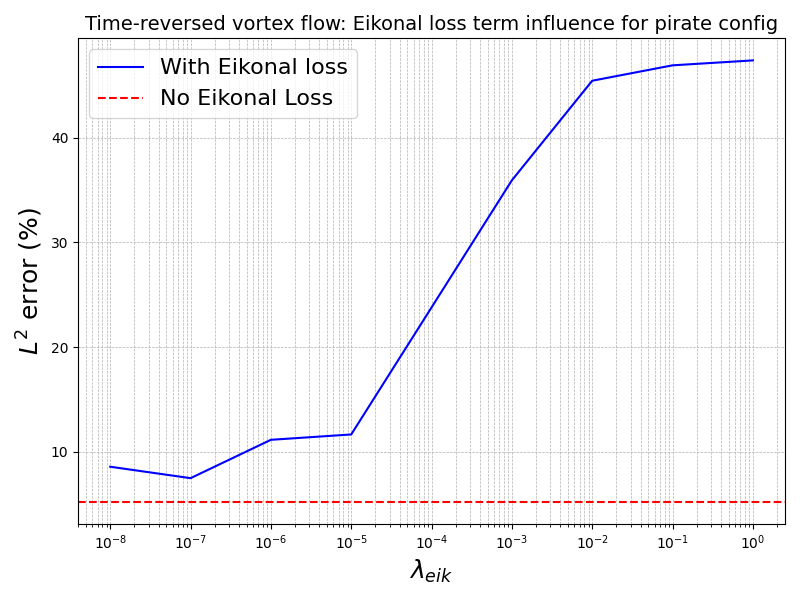}
        \caption{Influence of \(\lambda_{eik}\) on \textit{PirateNet}}
        \label{fig:vortex_eik_pirate}
    \end{subfigure}
    \caption{Level set vortex: influence of the Eikonal loss term \(\lambda_{eik}\).}
    \label{fig:vortex_eik_influence}
\end{figure}

We began with the addition of the Eikonal term to the loss function. As we found previously, adding the Eikonal loss to the gradient normalization weight update worsened the results. Hence, we implemented the Eikonal loss as a regularization term kept constant throughout training. Using \textit{Plain}, \textit{Default} and \textit{PirateNet} configurations, we tested a range of weights, \(\lambda_{eik}\), and measured their influence on the error. 
Consistent with our previous experiment, we found that adding an Eikonal loss term to the loss function offers no significant improvement to the solution. However, in certain cases as with \textit{Default} and a carefully chosen weight, the Eikonal loss treated as a regularization term can slightly improve the results. Figure \ref{fig:vortex_eik_influence} demonstrates those results. 

\begin{figure}
    \centering
    \includegraphics[width=0.9\linewidth]{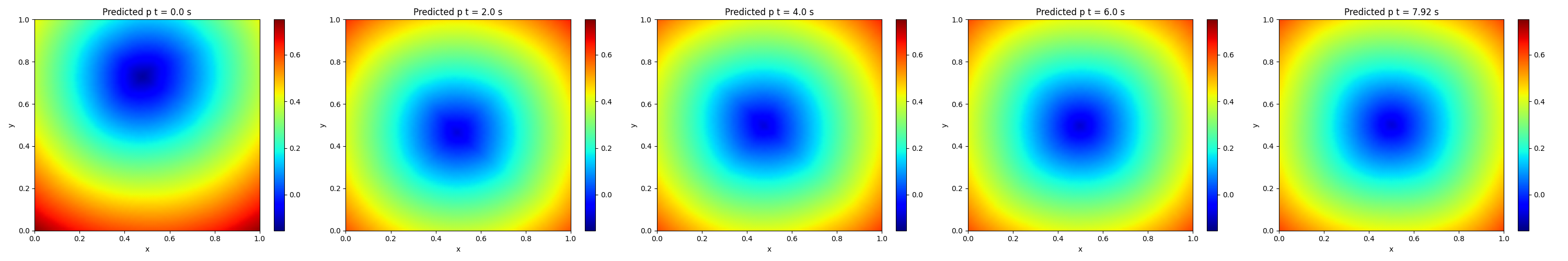}
    \caption{Level set vortex: Test Eikonal loss term.}
    \label{fig:levelset_eikonal}
\end{figure}

Plotting the predicted level set fields in Figure \ref{fig:levelset_eikonal} with \(\lambda_{eik} = 0.01\), we notice that the Eikonal term with a high weight tends to keep the interface in a circular shape. This prevents the interface from stretching and deforming as necessary with this problem.

We then tested the influence of adding a mass loss term. Starting with \textit{Method 1} from Section \ref{mass_loss}, we added the mass loss directly to the loss function using equation \ref{eq:ls_mass_loss_1}. Using the same hyperparameters as \textit{Sota}, we trained the network using a custom batch size of \(128 \times 128 \times 128\) spatio-temporal points for the mass loss term.
We tested with \(\lambda_{mass}\) in the gradient normalization or kept constant with varying values of \(\lambda_{mass}\).
Using \textit{Method 2}, we implemented the mass loss into a second network initialized with the trained weights of \textit{Sota}. For consistency, we kept the same batch size of \(128 \times 128 \times 128\) spatio-temporal points. The results are shown in Table \ref{tab:ls_vortex_l2_mass}.
\begin{table}[!h]
    \centering
        \begin{tabular}{c|cc}
            \hline
             Method& \(L^{2}\) error (\%) & MAPE (\%)\\
             \hline
             \textit{Sota} [no mass loss]& \textbf{0.85} &1.18\\
             \textit{Method 1} [Grad norm]&  1.00& 1.06\\
             \textit{Method 1} [\(\lambda_{mass}=0.1\)]&  1.08& 1.06\\
             \textit{Method 1} [\(\lambda_{mass}=1\)]&  1.02& 1.66\\
             \textit{Method 1} [\(\lambda_{mass}=10\)]&  1.02& 1.66\\
             \textit{Method 1} [\(\lambda_{mass}=100\)]&  1.00& 1.06\\
             \textit{Method 2}& 6.53& 16.8\\
        \end{tabular}
\caption{Level set vortex: influence of mass loss term.}
    \label{tab:ls_vortex_l2_mass}
\end{table}

\begin{figure}[!h]
    \centering
    \includegraphics[width=0.9\linewidth]{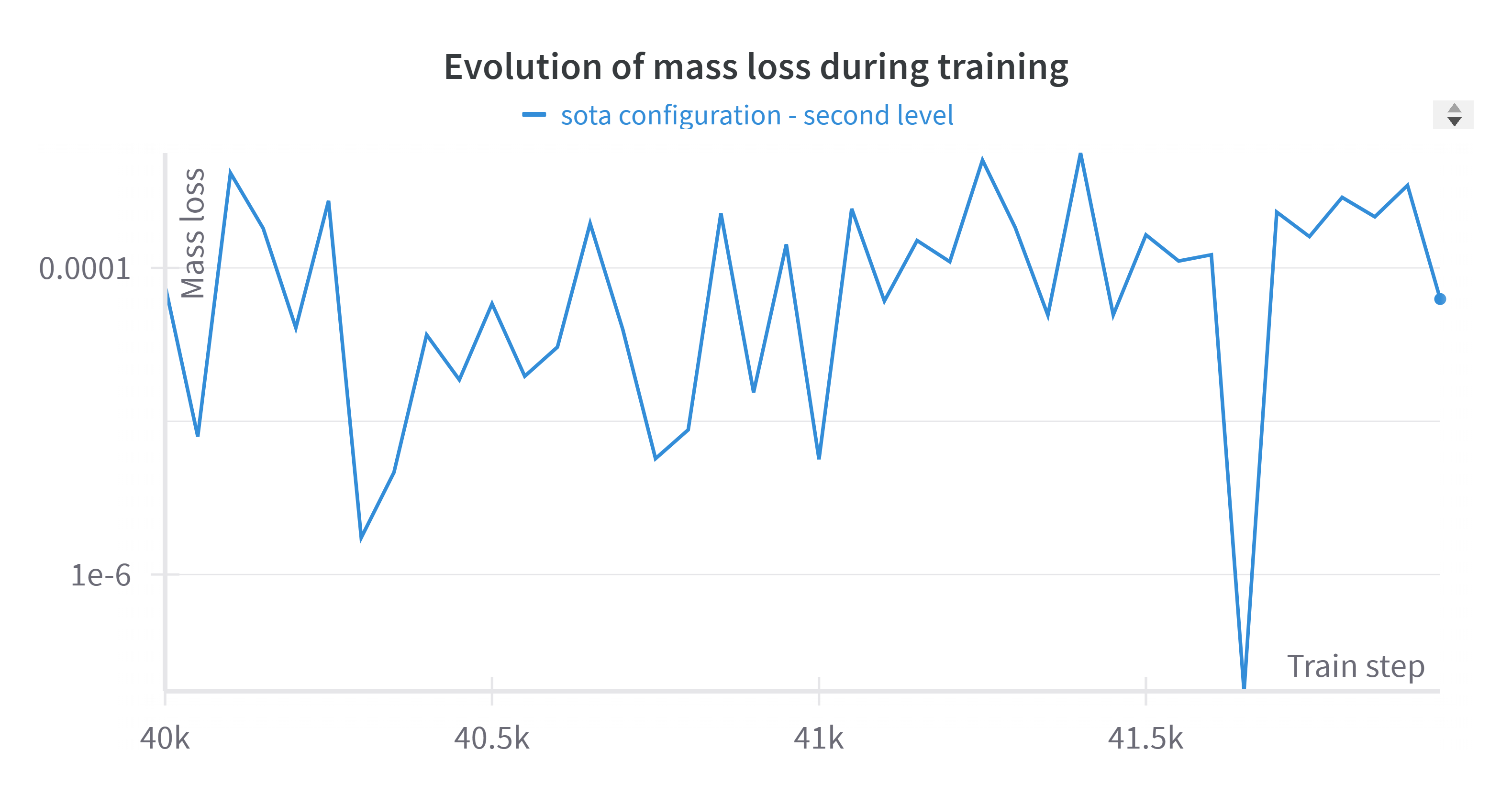}
    \caption{Level set vortex: Evolution of mass loss during training (2\textsuperscript{nd} level)}
    \label{fig:vortex_mass_loss}
\end{figure}

The benchmark \textit{Sota} without any mass loss terms remains the best configuration as neither method improves the \(L^2\) error or MAPE. Varying  \(\lambda_{mass}\), we see that the mass loss term has little influence on the accuracy of the model. Furthermore, from Figure \ref{fig:vortex_mass_loss}, we see with \textit{Method 2} that the mass loss fails to converge after 2000 iterations on the second level. From this, we can conclude that careful consideration should be taken before including the mass loss term in the loss function of level set-based problems.

This shows again that the \textit{PirateNet} architecture is capable of learning the solution to complex level set-based problems without adding Eikonal or mass loss terms.

\subsection{Coupled level set-Navier-Stokes equations: dam break flow application}
\label{levelset_dam}

Finally, we consider a very challenging simulation of a 2D dam break problem involving two immiscible, incompressible fluids using a PINN. The flow is governed by the incompressible Navier-Stokes equations coupled with a level set transport equation. The goal is to learn the temporal evolution of the level set, velocity and pressure fields and the interface between both fluids. The computational domain is fully enclosed (four-wall boundary condition), and the only body force acting is gravity.

Let $\Omega \subset \mathbb{R}^2$ be a bounded domain with coordinates $(x, y)$ and $t \in [0, T]$ the temporal domain. We define:

\begin{itemize}
\item $u(t, x, y), v(t, x, y)$: velocity components in the $x$ and $y$ directions
\item $p(t, x, y)$: pressure field
\item $\phi(t, x, y)$: level set function describing the interface between fluids
\item $\rho(\phi), \mu(\phi)$: fluid density and viscosity defined as functions of $\phi$
\item $\mathbf{g} = (0, -g)$: gravitational acceleration acting only in the vertical direction
\end{itemize}

The level set function is chosen such that:
\begin{itemize}
\item $\phi < 0$ in fluid 1 (dense fluid)
\item $\phi > 0$ in fluid 2 (light fluid)
\item $\phi = 0$ at the interface
\end{itemize}

Careful consideration needs to be taken to enforce the material properties of both fluids. To differentiate both fluids, we employ a smoothed Heaviside function $H_\epsilon(\phi)$:
\begin{equation}
H_\epsilon(\phi) = \frac{1}{2} \left(1 + \tanh\left(\frac{\phi}{\epsilon}\right)\right)
\label{eq:smoothed_heaviside}
\end{equation}

This gives smooth transitions for density and viscosity enabling stable gradient flow across the fluid interface:
\begin{align}
\rho(\phi) &= \rho_1 (1 - H_\epsilon(\phi)) + \rho_2 H_\epsilon(\phi) \\
\mu(\phi) &= \mu_1 (1 - H_\epsilon(\phi)) + \mu_2 H_\epsilon(\phi)
\end{align}

\subsubsection{Nondimensionalization of Coupled Level Set-Navier-Stokes Equations}

In CFD, dimensionless equations are preferred because they simplify equations by removing the dependence on specific units and making them universally applicable at different physical scales. This approach also improves the numerical stability of PINN training by scaling variables to similar orders of magnitude. Using dimensionless equations offers similar benefits as data scaling, a crucial step in traditional machine learning. The dimensionless version of the coupled level set-Navier-Stokes problem is hence used. Furthermore, curriculum training \cite{krishnapriyan_characterizing_2021} could then be easily used by gradually varying the nondimensional parameters governing the problem.

We define the following characteristic scales for the dam break problem:
\begin{itemize}
\item Characteristic Height: $H$ = initial column height of fluid 1
\item Characteristic Velocity: $U = \sqrt{g\,H}$ (free fall velocity)
\item Characteristic Time: $T = \sqrt{H/g}$ (free fall time)
\item Characteristic Pressure: $P = \rho_{\text{ref}} \, g \, H$ (hydrostatic pressure)
\end{itemize}
The density and viscosity of fluid 1 are used as reference values for the nondimensionalization.
The following nondimensional variables are set:
\paragraph{Spatial and Temporal Variables:}
\begin{align}
x^* = \frac{x}{H}, \quad y^* = \frac{y}{H}, \quad t^* = \frac{t}{T} = t\sqrt{\frac{g}{H}}
\end{align}

\paragraph{Field Variables:}
\begin{align}
u^* = \frac{u}{U} = \frac{u}{\sqrt{gH}}, \quad v^* = \frac{v}{U} = \frac{v}{\sqrt{gH}}
\end{align}
\begin{align}
p^* = \frac{p}{P} = \frac{p}{\rho_{\text{ref}} g H}, \quad \phi^* = \frac{\phi}{H}
\end{align}

\paragraph{Material Properties:}
\begin{align}
\rho^*(\phi) = \frac{\rho(\phi)}{\rho_{\text{ref}}}, \quad \mu^*(\phi) = \frac{\mu(\phi)}{\mu_{\text{ref}}}
\end{align}
Using these variables and the chain rule, we can derive the dimensionless equations. Below are the final forms used for training our PINNs.
\paragraph{Non-Dimensional Level Set Residual:}
\begin{align}
\mathcal{R}_{\phi} = \frac{\partial \phi^*}{\partial t^*} + u^*\frac{\partial \phi^*}{\partial x^*} + v^*\frac{\partial \phi^*}{\partial y^*}
\end{align}
\paragraph{Non-Dimensional Continuity Residual:}
\begin{align}
\mathcal{R}_{\text{cont}} = \frac{\partial u^*}{\partial x^*} + \frac{\partial v^*}{\partial y^*}
\end{align}
\paragraph{Non-Dimensional X-Momentum Residual:}
\begin{align}
\begin{aligned}
\mathcal{R}_{\text{mom-x}} &= \frac{\partial u^*}{\partial t^*} + u^*\frac{\partial u^*}{\partial x^*} + v^*\frac{\partial u^*}{\partial y^*} + \frac{1}{\rho^*(\phi)}\frac{\partial p^*}{\partial x^*} \\
&- \frac{1}{\text{Re}} \cdot \frac{\mu^*(\phi) }{\rho^*(\phi)}\left(\frac{\partial^2 u^*}{\partial (x^*)^2} + \frac{\partial^2 u^*}{\partial (y^*)^2}\right) - g^*_x
\end{aligned}
\end{align}
\paragraph{Non-Dimensional Y-Momentum Residual:}
\begin{align}
\begin{aligned}
\mathcal{R}_{\text{mom-y}} &= \frac{\partial v^*}{\partial t^*} + u^*\frac{\partial v^*}{\partial x^*} + v^*\frac{\partial v^*}{\partial y^*} + \frac{1}{\rho^*(\phi)}\frac{\partial p^*}{\partial y^*} \\
&- \frac{1}{\text{Re}} \cdot \frac{\mu^*(\phi) }{\rho^*(\phi)}\left(\frac{\partial^2 v^*}{\partial (x^*)^2} +  \frac{\partial^2 v^*}{\partial (y^*)^2}\right) - g^*_y
\end{aligned}
\end{align}

We note that to reduce computational complexity and enable more stable training, only the dominant second-order viscous terms are retained for the momentum equations. The terms involving derivation of $\mu$ are dropped under the assumption that the residuals aren't computed for collocation points exactly on the sharp interface. There is an implicit continuity of the total Cauchy stress across the interface.

\subsubsection{Problem setup}
The fluid properties for this problem are resumed in Table \ref{tab:ls_coupled_fluid_properties}. The Heaviside function \ref{eq:smoothed_heaviside} is smoothed with $\epsilon=0.01$ while gravity is set as $g=9.81 \frac{m}{s^2}$, acting in the negative $y$ direction. The spatial domain is defined as a square box of length $L=1.168 \,m$ on each side and the simulation is run for $0.8$ seconds. The denser Fluid 1 is initially contained in a rectangle of width $0.292 \,m$ and height $0.584 \,m$. Hence, the nondimensional constants are set as $H=0.584\,m$, $U=2.4\, \frac{m}{s}$, $T=0.24\,s$ and $P=5729 \,Pa$. Using these values and using the fluid properties of the denser fluid, the reference Reynolds number is computed as $Re=\frac{\rho_{1}*\sqrt{g*H}*H}{mu_1}=35*10^3$.
\begin{table}[!h]
    \centering
    \resizebox{0.8\textwidth}{!}{
        \begin{tabular}{c|cc}
            \hline
             Fluid& Density $\rho$ ($\frac{kg}{m^3}$)& Dynamic viscosity $\mu$ ($Pa*s$)\\
             \hline
             Fluid 1 (dense)& 1000&0.04\\
             Fluid 2 (light)&  50& 0.02\\
        \end{tabular}
    }
\caption{Coupled level set-NS: fluid properties.}
    \label{tab:ls_coupled_fluid_properties}
\end{table}
The reference solution was generated using the same FEM solver as previous sections. The results were interpolated on a $41\times41$ grid for $160$ time steps.

\subsubsection{Geometric reinitialization}
As with classical level set solvers based on the FEM, the signed distance property of the level set function is gradually lost during the training of our PINN. Although regularization with an Eikonal loss term helps enforce the signed distance property to some extent, we observed that it was insufficient to maintain stability over long time horizons.

To address this, we propose a simple geometric reinitialization step that is applied between time windows of sequence-to-sequence training. Before training begins on the new window of sequence-to-sequence, we reinitialize the level set field to restore its signed distance property. The geometric reinitialization method is based on the Euclidean distance transform (EDT), a function already available in Scipy \cite{virtanen_scipy_2020}. This method produces a new level set field where the zero level set remains in place, but the values on either side are reset to their signed distances from the interface.

This approach was found to significantly improve the stability and convergence of training for the coupled Navier-Stokes and level set PINN solver, especially when dealing with high density ratios and sharp interface dynamics.

\subsubsection{Results}
\begin{figure}
    \centering
    \includegraphics[width=0.9\linewidth]{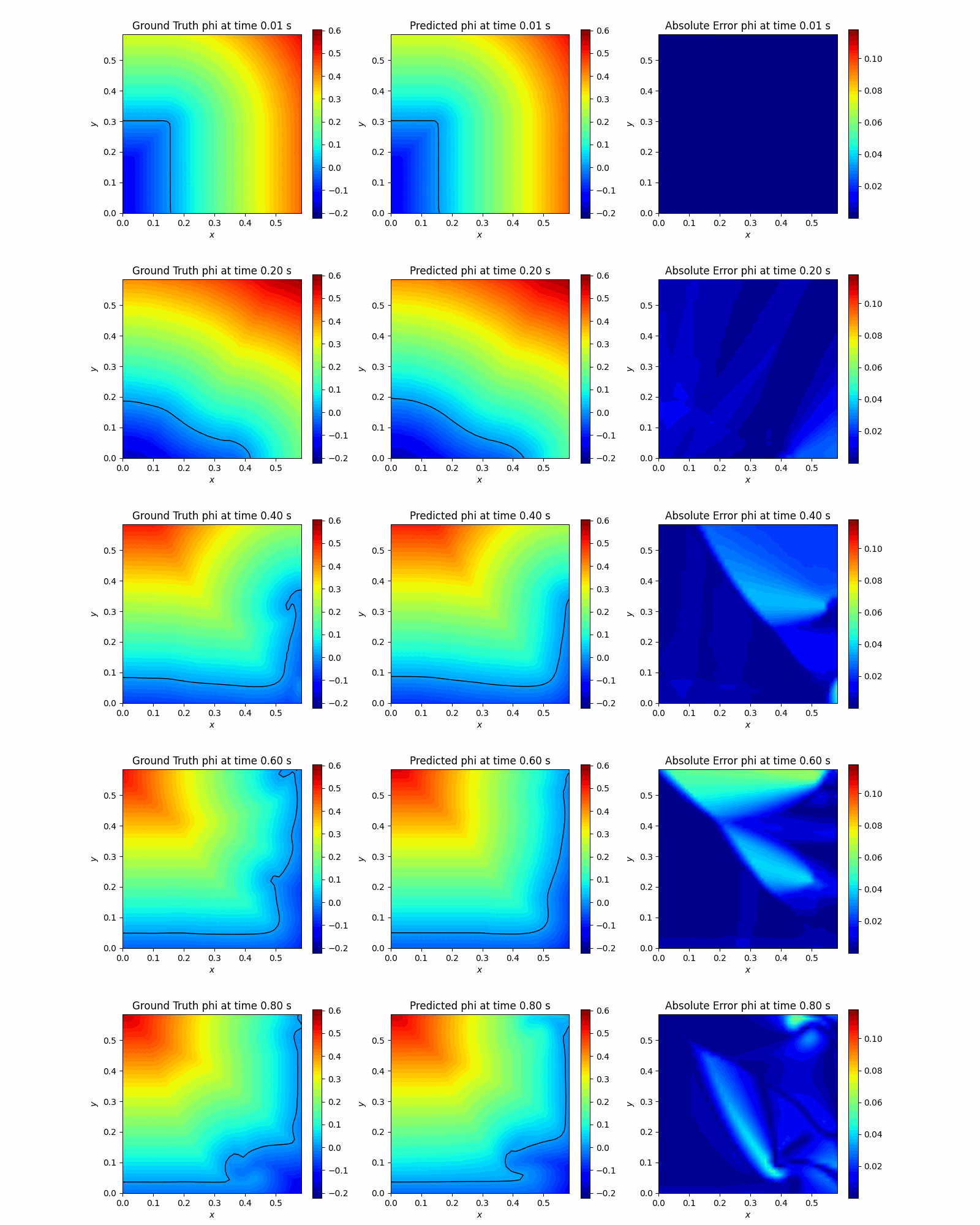}
    \caption{Coupled level set-NS: Temporal evolution of $\phi$ for the FEM reference and PINN}
    \label{fig:ls_coupled_evol}
\end{figure}

Given the computational cost of this problem, a Modified MLP was used instead of the PirateNet. Generally the cost of PirateNet is about three times that of the Modified MLP. This allows to fit more collocation points in the memory of a single V100 GPU. 
The SOAP optimizer is used due to the complex nature of this coupled problem with several conflicting loss terms, three inputs and four outputs. A more detailed list of hyperparameters can be found in Table \ref{tab:ls_coupled_hp}.
With the proposed method and hyperparameters, the PINN obtained a relative $L^2$ error when compared to the FEM reference of $L^2=5.6\%$. We note that the frames considered in this error calculation are only the ones subjected to the reinitialization step. Figure \ref{fig:ls_coupled_evol} shows the evolution of the reference and predicted level set fields.

It can be seen that the PINN solver generally captures the evolution of the level set field but struggles with the sharper details such as splashes. This error could probably be lowered with more collocation points and time windows, but this would significantly increase computational cost.

\begin{figure}
    \centering
    \includegraphics[width=0.6\linewidth]{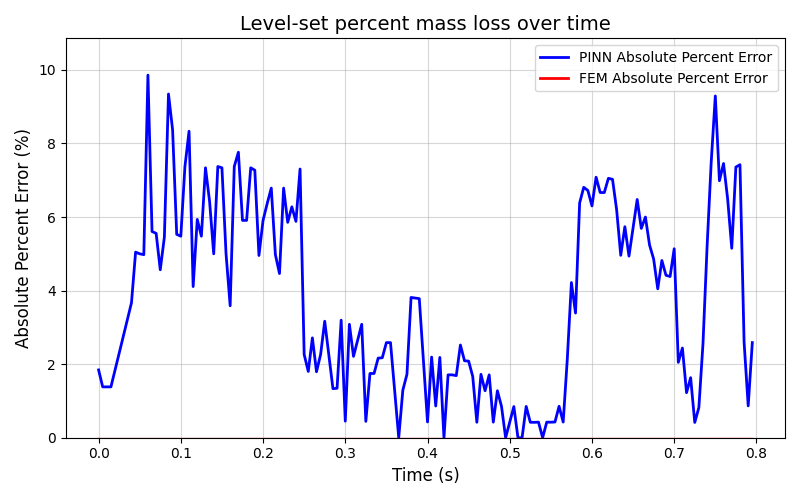}
    \caption{Coupled level set-NS: Temporal evolution of mass loss}
    \label{fig:ls_coupled_mass}
\end{figure}

Furthermore, the mean absolute error of mass loss is $Mass_{MAPE}=3.8\%$. The evolution of mass loss plotted in Figure \ref{fig:ls_coupled_mass} shows that the PINN suffers from large mass loss after the sudden drop of the dense fluid column, but is capable of restoring it's mass in time. A significant pattern of error accumulation isn't noticed.
A significant downfall of solving this problem with a PINN is the computational requirements. The PINN took 16 hours to train on a single V100 GPU while the reference FEM took 1.25 hours on a laptop CPU.

Nevertheless, this example shows that PINNs can capture the evolution of a level set field for a complex coupled level set-Navier-Stokes problem.

%%%%%%%%%%%%%%%%%%%%%%%%%%

\section{Conclusion}
\label{Conclusion}

This study has demonstrated the efficacy of advanced PINN architectures such as \textit{PirateNet} and appropriate training strategies in resolving complex level set problems, particularly those characterized by rigid-body motion and significant interface deformation. Across a series of numerical benchmarks, including Burgers' equations, Zalesak's disk, and the time-reversed vortex flow, \textit{PirateNet} consistently surpassed conventional PINN frameworks in terms of accuracy and computational stability. A PINN with the Modified MLP architecture also successfully solved a complex dam break problem in two phase flow by coupling the level set equation with the Navier-Stokes equations.

Including Eikonal and mass conservation loss terms within the training process yielded negligible effects across the tested benchmarks. Although the Eikonal term may serve as a viable regularization component in certain contexts, our findings suggest that it detrimentally affects problems involving pronounced interface deformation if the weight is not well chosen. For the benchmarks tested, we found that the weight, \(\lambda_{eik}\), should be well chosen to avoid completely altering the solution. Similarly, mass conservation terms, despite their conceptual appeal, introduced prohibitive computational overhead without delivering meaningful improvements in solution accuracy.

These results highlight that the core enhancements intrinsic to \textit{PirateNet}, such as causal training, sequence-to-sequence learning, random weight factorization, and Fourier feature embeddings, are sufficiently robust for a wide range of level set applications. These enhancements allow PINNs to learn the level set equations without the need for upwind numerical stabilization or mass conservation schemes as used in classic numerical methods. However, in the case of more complex problems like the tested dam break where the signed distance property needs to be reinitialized, a geometric reinitialization can be introduced between time windows of sequence-to-sequence training to allow long time inference.

In less complex scenarios, such as the 1D and 2D Burgers’ equations, the performance differential between \textit{PirateNet} and alternative architectures was less pronounced, as all configurations achieved comparably low error rates. However, the utility of \textit{PirateNet} became unequivocally evident in tackling more sophisticated challenges. For instance, in the time-reversed vortex flow benchmark, \textit{PirateNet} exhibited a fourfold reduction in error relative to an improved PINN and a tenfold reduction compared to the original PINN. With optimized hyperparameters, the architecture achieved a remarkably low error rate of \(L^2 = 0.85\%\). However, \textit{PirateNet} requires more memory than alternatives like the Modified MLP which makes \textit{PirateNet} harder to use in limited memory application.

Nevertheless, gaps persist between the capabilities of PINNs and traditional finite element methods (FEM), particularly with respect to mass conservation in scenarios involving significant interface deformation. The FEM solver showed better mass conservation, with a mean absolute percent error of 0.07\% compared to 1.18\% for \textit{PirateNet} in the time-reversed vortex flow benchmark.

Future research should prioritize addressing these limitations. Hybrid methodologies that integrate FEM strengths with PINN flexibility or innovations in loss function design explicitly targeting mass conservation could provide promising solutions. Additionally, extending \textit{PirateNet}’s framework to encompass real-world applications with heightened physical complexity represents an exciting path for future development.

In summary, this study constitutes a substantial advancement in the application of the physics-informed neural networks approach by extending \textit{PirateNet}'s architecture to accurately model dynamic interfaces without reliance on upwind numerical stabilization or explicit mass conservation schemes. However, further investigation is imperative to fully bridge the performance gap with traditional numerical solvers and unlock its full potential for both academic inquiry and practical deployment.

\subsection*{\bf Acknowledgments}

This research was supported by the Natural Sciences and Engineering
Research Council of Canada (Grant number RGPIN/2693-2021). The financial support is gratefully acknowledged.

\subsection*{\bf Code availability}
 The software can be shared upon reasonable
request.

\subsection*{\bf Declarations of Competing Interest}
The authors declare that they have
no known competing financial interests or personal relationships that could have appeared to influence the work reported in this paper.

\subsection*{\bf Contribution Statement} Mathieu Mullins: Methodology, Software,
Validation, Writing. 
Hamza Kamil: Methodology, Software,
Validation, review.
Adil Fahsi: FEM code, validation, review.
Azzeddine Soulaïmani: Methodology, Validation,
 review, Project conceptualization,  funding acquisition,
 Supervision.

\bibliographystyle{elsarticle-num}

\bibliography{references.bib}

\begin{thebibliography}{10}
\expandafter\ifx\csname url\endcsname\relax
  \def\url#1{\texttt{#1}}\fi
\expandafter\ifx\csname urlprefix\endcsname\relax\def\urlprefix{URL }\fi
\expandafter\ifx\csname href\endcsname\relax
  \def\href#1#2{#2} \def\path#1{#1}\fi

\bibitem{cheng_improved_2024}
S.~Cheng, M.~Yang, C.~Li, H.~Xu, C.~Chen, D.~Shu, Y.~Jiang, Y.~Gui, N.~Dong, \href{https://doi.org/10.1007/s11269-024-03914-9}{An {Improved} {Coupled} {Hydrologic}-{Hydrodynamic} {Model} for {Urban} {Flood} {Simulations} {Under} {Varied} {Scenarios}}, Water Resources Management (Aug. 2024).
\newblock \href {https://doi.org/10.1007/s11269-024-03914-9} {\path{doi:10.1007/s11269-024-03914-9}}.
\newline\urlprefix\url{https://doi.org/10.1007/s11269-024-03914-9}

\bibitem{google_research_flood_2023}
G.~Research, \href{https://sites.research.google/floodforecasting/}{Flood {Forecasting} - {Google} {Research}} (2023).
\newline\urlprefix\url{https://sites.research.google/floodforecasting/}

\bibitem{harlow_numerical_1965}
F.~H. Harlow, J.~E. Welch, \href{https://doi.org/10.1063/1.1761178}{Numerical {Calculation} of {Time}‐{Dependent} {Viscous} {Incompressible} {Flow} of {Fluid} with {Free} {Surface}}, The Physics of Fluids 8~(12) (1965) 2182--2189.
\newblock \href {https://doi.org/10.1063/1.1761178} {\path{doi:10.1063/1.1761178}}.
\newline\urlprefix\url{https://doi.org/10.1063/1.1761178}

\bibitem{unverdi_front-tracking_1992}
S.~O. Unverdi, G.~Tryggvason, \href{https://www.sciencedirect.com/science/article/pii/002199919290307K}{A front-tracking method for viscous, incompressible, multi-fluid flows}, Journal of Computational Physics 100~(1) (1992) 25--37.
\newblock \href {https://doi.org/10.1016/0021-9991(92)90307-K} {\path{doi:10.1016/0021-9991(92)90307-K}}.
\newline\urlprefix\url{https://www.sciencedirect.com/science/article/pii/002199919290307K}

\bibitem{noh_slic_1976}
W.~F. Noh, P.~Woodward, {SLIC} ({Simple} {Line} {Interface} {Calculation}), in: A.~I. van~de Vooren, P.~J. Zandbergen (Eds.), Proceedings of the {Fifth} {International} {Conference} on {Numerical} {Methods} in {Fluid} {Dynamics} {June} 28 – {July} 2, 1976 {Twente} {University}, {Enschede}, Springer, Berlin, Heidelberg, 1976, pp. 330--340.
\newblock \href {https://doi.org/10.1007/3-540-08004-X_336} {\path{doi:10.1007/3-540-08004-X_336}}.

\bibitem{osher_fronts_1988}
S.~Osher, J.~A. Sethian, \href{https://www.sciencedirect.com/science/article/pii/0021999188900022}{Fronts propagating with curvature-dependent speed: {Algorithms} based on {Hamilton}-{Jacobi} formulations}, Journal of Computational Physics 79~(1) (1988) 12--49.
\newblock \href {https://doi.org/10.1016/0021-9991(88)90002-2} {\path{doi:10.1016/0021-9991(88)90002-2}}.
\newline\urlprefix\url{https://www.sciencedirect.com/science/article/pii/0021999188900022}

\bibitem{malladi_shape_1995}
R.~Malladi, J.~Sethian, B.~Vemuri, Shape {Modeling} with {Front} {Propagation}: {A} {Level} {Set} {Approach}, IEEE Transactions on Pattern Analysis and Machine Intelligence 17~(2) (1995) 158--175.
\newblock \href {https://doi.org/10.1109/34.368173} {\path{doi:10.1109/34.368173}}.

\bibitem{wang_level_2003}
M.~Y. Wang, X.~Wang, D.~Guo, \href{https://www.sciencedirect.com/science/article/pii/S0045782502005595}{A level set method for structural topology optimization}, Computer Methods in Applied Mechanics and Engineering 192~(1) (2003) 227--246.
\newblock \href {https://doi.org/10.1016/S0045-7825(02)00559-5} {\path{doi:10.1016/S0045-7825(02)00559-5}}.
\newline\urlprefix\url{https://www.sciencedirect.com/science/article/pii/S0045782502005595}

\bibitem{gomes_reconciling_2000}
J.~Gomes, O.~Faugeras, \href{https://www.sciencedirect.com/science/article/pii/S104732039990439X}{Reconciling {Distance} {Functions} and {Level} {Sets}}, Journal of Visual Communication and Image Representation 11~(2) (2000) 209--223.
\newblock \href {https://doi.org/10.1006/jvci.1999.0439} {\path{doi:10.1006/jvci.1999.0439}}.
\newline\urlprefix\url{https://www.sciencedirect.com/science/article/pii/S104732039990439X}

\bibitem{toure_stabilized_2016}
M.~K. Touré, A.~Soulaïmani, \href{https://www.sciencedirect.com/science/article/pii/S0898122116300827}{Stabilized finite element methods for solving the level set equation without reinitialization}, Computers \& Mathematics with Applications 71~(8) (2016) 1602--1623.
\newblock \href {https://doi.org/10.1016/j.camwa.2016.02.028} {\path{doi:10.1016/j.camwa.2016.02.028}}.
\newline\urlprefix\url{https://www.sciencedirect.com/science/article/pii/S0898122116300827}

\bibitem{raissi_physics-informed_2019}
M.~Raissi, P.~Perdikaris, G.~E. Karniadakis, \href{https://www.sciencedirect.com/science/article/pii/S0021999118307125}{Physics-informed neural networks: {A} deep learning framework for solving forward and inverse problems involving nonlinear partial differential equations}, Journal of Computational Physics 378 (2019) 686--707.
\newblock \href {https://doi.org/10.1016/j.jcp.2018.10.045} {\path{doi:10.1016/j.jcp.2018.10.045}}.
\newline\urlprefix\url{https://www.sciencedirect.com/science/article/pii/S0021999118307125}

\bibitem{baydin_automatic_2017}
A.~G. Baydin, B.~A. Pearlmutter, A.~A. Radul, J.~M. Siskind, Automatic differentiation in machine learning: a survey, J. Mach. Learn. Res. 18~(1) (2017) 5595--5637.

\bibitem{fang_deep_2020}
Z.~Fang, J.~Zhan, \href{https://ieeexplore.ieee.org/document/8946546/?arnumber=8946546}{Deep {Physical} {Informed} {Neural} {Networks} for {Metamaterial} {Design}}, IEEE Access 8 (2020) 24506--24513, conference Name: IEEE Access.
\newblock \href {https://doi.org/10.1109/ACCESS.2019.2963375} {\path{doi:10.1109/ACCESS.2019.2963375}}.
\newline\urlprefix\url{https://ieeexplore.ieee.org/document/8946546/?arnumber=8946546}

\bibitem{goswami_physics-informed_2022}
S.~Goswami, M.~Yin, Y.~Yu, G.~Karniadakis, \href{http://arxiv.org/abs/2108.06905}{A physics-informed variational {DeepONet} for predicting the crack path in brittle materials}, Computer Methods in Applied Mechanics and Engineering 391 (2022) 114587, arXiv:2108.06905 [cs, math].
\newblock \href {https://doi.org/10.1016/j.cma.2022.114587} {\path{doi:10.1016/j.cma.2022.114587}}.
\newline\urlprefix\url{http://arxiv.org/abs/2108.06905}

\bibitem{liu_physics-informed_2021}
S.~Liu, B.~B. Kappes, B.~Amin-ahmadi, O.~Benafan, X.~Zhang, A.~P. Stebner, \href{https://www.sciencedirect.com/science/article/pii/S2352940720303462}{Physics-informed machine learning for composition – process – property design: {Shape} memory alloy demonstration}, Applied Materials Today 22 (2021) 100898.
\newblock \href {https://doi.org/10.1016/j.apmt.2020.100898} {\path{doi:10.1016/j.apmt.2020.100898}}.
\newline\urlprefix\url{https://www.sciencedirect.com/science/article/pii/S2352940720303462}

\bibitem{salvati_defect-based_2022}
E.~Salvati, A.~Tognan, L.~Laurenti, M.~Pelegatti, F.~De~Bona, \href{https://www.sciencedirect.com/science/article/pii/S0264127522007110}{A defect-based physics-informed machine learning framework for fatigue finite life prediction in additive manufacturing}, Materials \& Design 222 (2022) 111089.
\newblock \href {https://doi.org/10.1016/j.matdes.2022.111089} {\path{doi:10.1016/j.matdes.2022.111089}}.
\newline\urlprefix\url{https://www.sciencedirect.com/science/article/pii/S0264127522007110}

\bibitem{cho_lstm-pinn_2022}
G.~Cho, D.~Zhu, J.~J. Campbell, M.~Wang, \href{https://ieeexplore.ieee.org/abstract/document/9895422}{An {LSTM}-{PINN} {Hybrid} {Method} to {Estimate} {Lithium}-{Ion} {Battery} {Pack} {Temperature}}, IEEE Access 10 (2022) 100594--100604, conference Name: IEEE Access.
\newblock \href {https://doi.org/10.1109/ACCESS.2022.3208103} {\path{doi:10.1109/ACCESS.2022.3208103}}.
\newline\urlprefix\url{https://ieeexplore.ieee.org/abstract/document/9895422}

\bibitem{pombo_increasing_2022}
D.~V. Pombo, H.~W. Bindner, S.~V. Spataru, P.~E. Sørensen, P.~Bacher, \href{https://www.mdpi.com/1424-8220/22/3/749}{Increasing the {Accuracy} of {Hourly} {Multi}-{Output} {Solar} {Power} {Forecast} with {Physics}-{Informed} {Machine} {Learning}}, Sensors 22~(3) (2022) 749, number: 3 Publisher: Multidisciplinary Digital Publishing Institute.
\newblock \href {https://doi.org/10.3390/s22030749} {\path{doi:10.3390/s22030749}}.
\newline\urlprefix\url{https://www.mdpi.com/1424-8220/22/3/749}

\bibitem{liu_surrogate_2023}
K.~Liu, K.~Luo, Y.~Cheng, A.~Liu, H.~Li, J.~Fan, S.~Balachandar, \href{https://www.sciencedirect.com/science/article/pii/S0010218023004698}{Surrogate modeling of parameterized multi-dimensional premixed combustion with physics-informed neural networks for rapid exploration of design space}, Combustion and Flame 258 (2023) 113094.
\newblock \href {https://doi.org/10.1016/j.combustflame.2023.113094} {\path{doi:10.1016/j.combustflame.2023.113094}}.
\newline\urlprefix\url{https://www.sciencedirect.com/science/article/pii/S0010218023004698}

\bibitem{masclans_thermodynamics-informed_2023}
N.~Masclans, F.~Vázquez-Novoa, M.~Bernades, R.~M. Badia, L.~Jofre, \href{https://www.sciencedirect.com/science/article/pii/S2666202723001635}{Thermodynamics-informed neural network for recovering supercritical fluid thermophysical information from turbulent velocity data}, International Journal of Thermofluids 20 (2023) 100448.
\newblock \href {https://doi.org/10.1016/j.ijft.2023.100448} {\path{doi:10.1016/j.ijft.2023.100448}}.
\newline\urlprefix\url{https://www.sciencedirect.com/science/article/pii/S2666202723001635}

\bibitem{dazzi_physics-informed_2024}
S.~Dazzi, \href{https://onlinelibrary.wiley.com/doi/abs/10.1029/2023WR036589}{Physics-{Informed} {Neural} {Networks} for the {Augmented} {System} of {Shallow} {Water} {Equations} {With} {Topography}}, Water Resources Research 60~(10) (2024) e2023WR036589, \_eprint: https://onlinelibrary.wiley.com/doi/pdf/10.1029/2023WR036589.
\newblock \href {https://doi.org/10.1029/2023WR036589} {\path{doi:10.1029/2023WR036589}}.
\newline\urlprefix\url{https://onlinelibrary.wiley.com/doi/abs/10.1029/2023WR036589}

\bibitem{donnelly_physics-informed_2024}
J.~Donnelly, A.~Daneshkhah, S.~Abolfathi, \href{https://linkinghub.elsevier.com/retrieve/pii/S0048969723074430}{Physics-informed neural networks as surrogate models of hydrodynamic simulators}, Science of The Total Environment 912 (2024) 168814.
\newblock \href {https://doi.org/10.1016/j.scitotenv.2023.168814} {\path{doi:10.1016/j.scitotenv.2023.168814}}.
\newline\urlprefix\url{https://linkinghub.elsevier.com/retrieve/pii/S0048969723074430}

\bibitem{li_improved_2024}
Y.~Li, Q.~Sun, J.~Wei, C.~Huang, \href{https://www.mdpi.com/2073-8994/16/10/1376}{An {Improved} {PINN} {Algorithm} for {Shallow} {Water} {Equations} {Driven} by {Deep} {Learning}}, Symmetry 16~(10) (2024) 1376, number: 10 Publisher: Multidisciplinary Digital Publishing Institute.
\newblock \href {https://doi.org/10.3390/sym16101376} {\path{doi:10.3390/sym16101376}}.
\newline\urlprefix\url{https://www.mdpi.com/2073-8994/16/10/1376}

\bibitem{qi_physics-informed_2024}
X.~Qi, G.~A.~M. de~Almeida, S.~Maldonado, \href{https://www.sciencedirect.com/science/article/pii/S0022169424006589}{Physics-informed neural networks for solving flow problems modeled by the {2D} {Shallow} {Water} {Equations} without labeled data}, Journal of Hydrology 636 (2024) 131263.
\newblock \href {https://doi.org/10.1016/j.jhydrol.2024.131263} {\path{doi:10.1016/j.jhydrol.2024.131263}}.
\newline\urlprefix\url{https://www.sciencedirect.com/science/article/pii/S0022169424006589}

\bibitem{kamil_transfer_2024}
H.~Kamil, A.~Soulaïmani, A.~Beljadid, \href{https://www.sciencedirect.com/science/article/pii/S0045782524005322}{A transfer learning physics-informed deep learning framework for modeling multiple solute dynamics in unsaturated soils}, Computer Methods in Applied Mechanics and Engineering 431 (2024) 117276.
\newblock \href {https://doi.org/10.1016/j.cma.2024.117276} {\path{doi:10.1016/j.cma.2024.117276}}.
\newline\urlprefix\url{https://www.sciencedirect.com/science/article/pii/S0045782524005322}

\bibitem{rahaman_spectral_2019}
N.~Rahaman, A.~Baratin, D.~Arpit, F.~Draxler, M.~Lin, F.~Hamprecht, Y.~Bengio, A.~Courville, \href{https://proceedings.mlr.press/v97/rahaman19a.html}{On the {Spectral} {Bias} of {Neural} {Networks}}, in: Proceedings of the 36th {International} {Conference} on {Machine} {Learning}, PMLR, 2019, pp. 5301--5310, iSSN: 2640-3498.
\newline\urlprefix\url{https://proceedings.mlr.press/v97/rahaman19a.html}

\bibitem{wang_eigenvector_2021}
S.~Wang, H.~Wang, P.~Perdikaris, \href{https://www.sciencedirect.com/science/article/pii/S0045782521002759}{On the eigenvector bias of {Fourier} feature networks: {From} regression to solving multi-scale {PDEs} with physics-informed neural networks}, Computer Methods in Applied Mechanics and Engineering 384 (2021) 113938.
\newblock \href {https://doi.org/10.1016/j.cma.2021.113938} {\path{doi:10.1016/j.cma.2021.113938}}.
\newline\urlprefix\url{https://www.sciencedirect.com/science/article/pii/S0045782521002759}

\bibitem{wang_understanding_2021}
S.~Wang, Y.~Teng, P.~Perdikaris, \href{https://doi.org/10.1137/20M1318043}{Understanding and {Mitigating} {Gradient} {Flow} {Pathologies} in {Physics}-{Informed} {Neural} {Networks}}, SIAM J. Sci. Comput. 43~(5) (2021) A3055--A3081.
\newblock \href {https://doi.org/10.1137/20M1318043} {\path{doi:10.1137/20M1318043}}.
\newline\urlprefix\url{https://doi.org/10.1137/20M1318043}

\bibitem{wang_when_2022}
S.~Wang, X.~Yu, P.~Perdikaris, \href{https://www.sciencedirect.com/science/article/pii/S002199912100663X}{When and why {PINNs} fail to train: {A} neural tangent kernel perspective}, Journal of Computational Physics 449 (2022) 110768.
\newblock \href {https://doi.org/10.1016/j.jcp.2021.110768} {\path{doi:10.1016/j.jcp.2021.110768}}.
\newline\urlprefix\url{https://www.sciencedirect.com/science/article/pii/S002199912100663X}

\bibitem{wang_respecting_2022}
S.~Wang, S.~Sankaran, P.~Perdikaris, \href{https://arxiv.org/abs/2203.07404v1}{Respecting causality is all you need for training physics-informed neural networks} (Mar. 2022).
\newline\urlprefix\url{https://arxiv.org/abs/2203.07404v1}

\bibitem{tancik_fourier_2020}
M.~Tancik, P.~Srinivasan, B.~Mildenhall, S.~Fridovich-Keil, N.~Raghavan, U.~Singhal, R.~Ramamoorthi, J.~Barron, R.~Ng, \href{https://proceedings.neurips.cc/paper/2020/hash/55053683268957697aa39fba6f231c68-Abstract.html}{Fourier {Features} {Let} {Networks} {Learn} {High} {Frequency} {Functions} in {Low} {Dimensional} {Domains}}, in: Advances in {Neural} {Information} {Processing} {Systems}, Vol.~33, Curran Associates, Inc., 2020, pp. 7537--7547.
\newline\urlprefix\url{https://proceedings.neurips.cc/paper/2020/hash/55053683268957697aa39fba6f231c68-Abstract.html}

\bibitem{wang_random_2022}
S.~Wang, H.~Wang, J.~H. Seidman, P.~Perdikaris, \href{http://arxiv.org/abs/2210.01274}{Random {Weight} {Factorization} {Improves} the {Training} of {Continuous} {Neural} {Representations}}, arXiv:2210.01274 [cs] (Oct. 2022).
\newblock \href {https://doi.org/10.48550/arXiv.2210.01274} {\path{doi:10.48550/arXiv.2210.01274}}.
\newline\urlprefix\url{http://arxiv.org/abs/2210.01274}

\bibitem{daw_mitigating_2023}
A.~Daw, J.~Bu, S.~Wang, P.~Perdikaris, A.~Karpatne, \href{http://arxiv.org/abs/2207.02338}{Mitigating {Propagation} {Failures} in {Physics}-informed {Neural} {Networks} using {Retain}-{Resample}-{Release} ({R3}) {Sampling}}, arXiv:2207.02338 (Jun. 2023).
\newblock \href {https://doi.org/10.48550/arXiv.2207.02338} {\path{doi:10.48550/arXiv.2207.02338}}.
\newline\urlprefix\url{http://arxiv.org/abs/2207.02338}

\bibitem{wu_comprehensive_2023}
C.~Wu, M.~Zhu, Q.~Tan, Y.~Kartha, L.~Lu, \href{https://www.sciencedirect.com/science/article/pii/S0045782522006260}{A comprehensive study of non-adaptive and residual-based adaptive sampling for physics-informed neural networks}, Computer Methods in Applied Mechanics and Engineering 403 (2023) 115671.
\newblock \href {https://doi.org/10.1016/j.cma.2022.115671} {\path{doi:10.1016/j.cma.2022.115671}}.
\newline\urlprefix\url{https://www.sciencedirect.com/science/article/pii/S0045782522006260}

\bibitem{chen_gradnorm_2018}
Z.~Chen, V.~Badrinarayanan, C.-Y. Lee, A.~Rabinovich, \href{http://arxiv.org/abs/1711.02257}{{GradNorm}: {Gradient} {Normalization} for {Adaptive} {Loss} {Balancing} in {Deep} {Multitask} {Networks}}, arXiv:1711.02257 (Jun. 2018).
\newblock \href {https://doi.org/10.48550/arXiv.1711.02257} {\path{doi:10.48550/arXiv.1711.02257}}.
\newline\urlprefix\url{http://arxiv.org/abs/1711.02257}

\bibitem{bengio_curriculum_2009}
Y.~Bengio, J.~Louradour, R.~Collobert, J.~Weston, \href{https://dl.acm.org/doi/10.1145/1553374.1553380}{Curriculum learning}, in: Proceedings of the 26th {Annual} {International} {Conference} on {Machine} {Learning}, {ICML} '09, Association for Computing Machinery, New York, NY, USA, 2009, pp. 41--48.
\newblock \href {https://doi.org/10.1145/1553374.1553380} {\path{doi:10.1145/1553374.1553380}}.
\newline\urlprefix\url{https://dl.acm.org/doi/10.1145/1553374.1553380}

\bibitem{krishnapriyan_characterizing_2021}
A.~S. Krishnapriyan, A.~Gholami, S.~Zhe, R.~M. Kirby, M.~W. Mahoney, \href{http://arxiv.org/abs/2109.01050}{Characterizing possible failure modes in physics-informed neural networks}, arXiv:2109.01050 [physics] (Nov. 2021).
\newblock \href {https://doi.org/10.48550/arXiv.2109.01050} {\path{doi:10.48550/arXiv.2109.01050}}.
\newline\urlprefix\url{http://arxiv.org/abs/2109.01050}

\bibitem{wang_experts_2023}
S.~Wang, S.~Sankaran, H.~Wang, P.~Perdikaris, \href{http://arxiv.org/abs/2308.08468}{An {Expert}'s {Guide} to {Training} {Physics}-informed {Neural} {Networks}}, arXiv:2308.08468 [physics] (Aug. 2023).
\newblock \href {https://doi.org/10.48550/arXiv.2308.08468} {\path{doi:10.48550/arXiv.2308.08468}}.
\newline\urlprefix\url{http://arxiv.org/abs/2308.08468}

\bibitem{wang_piratenets_2025}
S.~Wang, B.~Li, Y.~Chen, P.~Perdikaris, \href{http://arxiv.org/abs/2402.00326}{{PirateNets}: {Physics}-informed {Deep} {Learning} with {Residual} {Adaptive} {Networks}}, Journal of Machine Learning Research 25 (2025) 1--51.
\newline\urlprefix\url{http://arxiv.org/abs/2402.00326}

\bibitem{m_silva_pinn-based_2024}
R.~M.~Silva, M.~Grave, A.~L. G.~A. Coutinho, \href{https://doi.org/10.1007/s00419-024-02622-5}{A {PINN}-based level-set formulation for reconstruction of bubble dynamics}, Archive of Applied Mechanics 94~(9) (2024) 2667--2682.
\newblock \href {https://doi.org/10.1007/s00419-024-02622-5} {\path{doi:10.1007/s00419-024-02622-5}}.
\newline\urlprefix\url{https://doi.org/10.1007/s00419-024-02622-5}

\bibitem{zhou_self-adaptive_2024}
W.~Zhou, S.~Miwa, K.~Okamoto, \href{https://doi.org/10.1063/5.0214646}{Self-adaptive and time divide-and-conquer physics-informed neural networks for two-phase flow simulations using interface tracking methods}, Physics of Fluids 36~(7) (2024) 073305.
\newblock \href {https://doi.org/10.1063/5.0214646} {\path{doi:10.1063/5.0214646}}.
\newline\urlprefix\url{https://doi.org/10.1063/5.0214646}

\bibitem{tang_physics-informed_2024}
M.~Tang, Z.~Xin, L.~Wang, \href{https://www.sciencedirect.com/science/article/pii/S0142727X2400376X}{Physics-{Informed} neural network for level set method in vapor condensation}, International Journal of Heat and Fluid Flow 110 (2024) 109651.
\newblock \href {https://doi.org/10.1016/j.ijheatfluidflow.2024.109651} {\path{doi:10.1016/j.ijheatfluidflow.2024.109651}}.
\newline\urlprefix\url{https://www.sciencedirect.com/science/article/pii/S0142727X2400376X}

\bibitem{zalesak_fully_1979}
S.~T. Zalesak, \href{https://www.sciencedirect.com/science/article/pii/0021999179900512}{Fully multidimensional flux-corrected transport algorithms for fluids}, Journal of Computational Physics 31~(3) (1979) 335--362.
\newblock \href {https://doi.org/10.1016/0021-9991(79)90051-2} {\path{doi:10.1016/0021-9991(79)90051-2}}.
\newline\urlprefix\url{https://www.sciencedirect.com/science/article/pii/0021999179900512}

\bibitem{sussman_level_1994}
M.~Sussman, P.~Smereka, S.~Osher, \href{https://www.sciencedirect.com/science/article/pii/S0021999184711557}{A {Level} {Set} {Approach} for {Computing} {Solutions} to {Incompressible} {Two}-{Phase} {Flow}}, Journal of Computational Physics 114~(1) (1994) 146--159.
\newblock \href {https://doi.org/10.1006/jcph.1994.1155} {\path{doi:10.1006/jcph.1994.1155}}.
\newline\urlprefix\url{https://www.sciencedirect.com/science/article/pii/S0021999184711557}

\bibitem{fahsi_numerical_2017}
A.~Fahsi, A.~Soulaïmani, \href{https://doi.org/10.1080/10618562.2017.1322200}{Numerical investigations of the {XFEM} for solving two-phase incompressible flows}, International Journal of Computational Fluid Dynamics 31~(3) (2017) 135--155, publisher: IAHR Website \_eprint: https://doi.org/10.1080/10618562.2017.1322200.
\newblock \href {https://doi.org/10.1080/10618562.2017.1322200} {\path{doi:10.1080/10618562.2017.1322200}}.
\newline\urlprefix\url{https://doi.org/10.1080/10618562.2017.1322200}

\bibitem{ausas_geometric_2011}
R.~F. Ausas, E.~A. Dari, G.~C. Buscaglia, \href{https://www.researchgate.net/publication/227728289_A_geometric_mass-preserving_redistancing_scheme_for_the_level_set_function}{A geometric mass‐preserving redistancing scheme for the level set function {\textbar} {Request} {PDF}}, International Journal for Numerical Methods in Fluids 65~(8) (2011) 989--1010.
\newblock \href {https://doi.org/10.1002/fld.2227} {\path{doi:10.1002/fld.2227}}.
\newline\urlprefix\url{https://www.researchgate.net/publication/227728289_A_geometric_mass-preserving_redistancing_scheme_for_the_level_set_function}

\bibitem{sethian_fast_1996}
J.~A. Sethian, \href{https://www.pnas.org/doi/abs/10.1073/pnas.93.4.1591}{A fast marching level set method for monotonically advancing fronts}, Proceedings of the National Academy of Sciences 93~(4) (1996) 1591--1595, publisher: Proceedings of the National Academy of Sciences.
\newblock \href {https://doi.org/10.1073/pnas.93.4.1591} {\path{doi:10.1073/pnas.93.4.1591}}.
\newline\urlprefix\url{https://www.pnas.org/doi/abs/10.1073/pnas.93.4.1591}

\bibitem{sethian_theory_1996}
J.~Sethian, Theory, algorithms, and applications of level set methods for propagating interfaces, Acta Numerica 5 (1996) 309--395.
\newblock \href {https://doi.org/10.1017/S0962492900002671} {\path{doi:10.1017/S0962492900002671}}.

\bibitem{li_level_2005}
C.~Li, C.~Xu, C.~Gui, M.~Fox, \href{https://ieeexplore.ieee.org/abstract/document/1467299?casa_token=ocTqkm535fwAAAAA:K54OoQY15TTx6g98got5ekiT0vvjjGdpG0W5JMAV12KarMT-J_CrK-5yQb6EGJEOtaqvg4JsLA}{Level set evolution without re-initialization: a new variational formulation}, in: 2005 {IEEE} {Computer} {Society} {Conference} on {Computer} {Vision} and {Pattern} {Recognition} ({CVPR}'05), Vol.~1, 2005, pp. 430--436 vol. 1, iSSN: 1063-6919.
\newblock \href {https://doi.org/10.1109/CVPR.2005.213} {\path{doi:10.1109/CVPR.2005.213}}.
\newline\urlprefix\url{https://ieeexplore.ieee.org/abstract/document/1467299?casa_token=ocTqkm535fwAAAAA:K54OoQY15TTx6g98got5ekiT0vvjjGdpG0W5JMAV12KarMT-J_CrK-5yQb6EGJEOtaqvg4JsLA}

\bibitem{van_der_pijl_mass-conserving_2005}
S.~P. van~der Pijl, A.~Segal, C.~Vuik, P.~Wesseling, \href{https://onlinelibrary.wiley.com/doi/abs/10.1002/fld.817}{A mass-conserving {Level}-{Set} method for modelling of multi-phase flows}, International Journal for Numerical Methods in Fluids 47~(4) (2005) 339--361, \_eprint: https://onlinelibrary.wiley.com/doi/pdf/10.1002/fld.817.
\newblock \href {https://doi.org/10.1002/fld.817} {\path{doi:10.1002/fld.817}}.
\newline\urlprefix\url{https://onlinelibrary.wiley.com/doi/abs/10.1002/fld.817}

\bibitem{olsson_conservative_2005}
E.~Olsson, G.~Kreiss, \href{https://www.sciencedirect.com/science/article/pii/S0021999105002184}{A conservative level set method for two phase flow}, Journal of Computational Physics 210~(1) (2005) 225--246.
\newblock \href {https://doi.org/10.1016/j.jcp.2005.04.007} {\path{doi:10.1016/j.jcp.2005.04.007}}.
\newline\urlprefix\url{https://www.sciencedirect.com/science/article/pii/S0021999105002184}

\bibitem{olsson_conservative_2007}
E.~Olsson, G.~Kreiss, S.~Zahedi, \href{https://www.sciencedirect.com/science/article/pii/S0021999107000046}{A conservative level set method for two phase flow {II}}, Journal of Computational Physics 225~(1) (2007) 785--807.
\newblock \href {https://doi.org/10.1016/j.jcp.2006.12.027} {\path{doi:10.1016/j.jcp.2006.12.027}}.
\newline\urlprefix\url{https://www.sciencedirect.com/science/article/pii/S0021999107000046}

\bibitem{penwarden_unified_2023}
M.~Penwarden, A.~D. Jagtap, S.~Zhe, G.~E. Karniadakis, R.~M. Kirby, \href{http://arxiv.org/abs/2302.14227}{A unified scalable framework for causal sweeping strategies for {Physics}-{Informed} {Neural} {Networks} ({PINNs}) and their temporal decompositions}, Journal of Computational Physics 493 (2023) 112464, arXiv:2302.14227 [physics].
\newblock \href {https://doi.org/10.1016/j.jcp.2023.112464} {\path{doi:10.1016/j.jcp.2023.112464}}.
\newline\urlprefix\url{http://arxiv.org/abs/2302.14227}

\bibitem{glorot_understanding_2010}
X.~Glorot, Y.~Bengio, \href{https://proceedings.mlr.press/v9/glorot10a.html}{Understanding the difficulty of training deep feedforward neural networks}, in: Proceedings of the {Thirteenth} {International} {Conference} on {Artificial} {Intelligence} and {Statistics}, JMLR Workshop and Conference Proceedings, 2010, pp. 249--256, iSSN: 1938-7228.
\newline\urlprefix\url{https://proceedings.mlr.press/v9/glorot10a.html}

\bibitem{mattey_novel_2022}
R.~Mattey, S.~Ghosh, \href{https://www.sciencedirect.com/science/article/pii/S0045782521006939}{A novel sequential method to train physics informed neural networks for {Allen} {Cahn} and {Cahn} {Hilliard} equations}, Computer Methods in Applied Mechanics and Engineering 390 (2022) 114474.
\newblock \href {https://doi.org/10.1016/j.cma.2021.114474} {\path{doi:10.1016/j.cma.2021.114474}}.
\newline\urlprefix\url{https://www.sciencedirect.com/science/article/pii/S0045782521006939}

\bibitem{wang_gradient_2025}
S.~Wang, A.~K. Bhartari, B.~Li, P.~Perdikaris, \href{http://arxiv.org/abs/2502.00604}{Gradient {Alignment} in {Physics}-informed {Neural} {Networks}: {A} {Second}-{Order} {Optimization} {Perspective}}, arXiv:2502.00604 [cs] (Feb. 2025).
\newblock \href {https://doi.org/10.48550/arXiv.2502.00604} {\path{doi:10.48550/arXiv.2502.00604}}.
\newline\urlprefix\url{http://arxiv.org/abs/2502.00604}

\bibitem{vyas_soap_2025}
N.~Vyas, D.~Morwani, R.~Zhao, M.~Kwun, I.~Shapira, D.~Brandfonbrener, L.~Janson, S.~Kakade, \href{http://arxiv.org/abs/2409.11321}{{SOAP}: {Improving} and {Stabilizing} {Shampoo} using {Adam}}, arXiv:2409.11321 [cs] (Feb. 2025).
\newblock \href {https://doi.org/10.48550/arXiv.2409.11321} {\path{doi:10.48550/arXiv.2409.11321}}.
\newline\urlprefix\url{http://arxiv.org/abs/2409.11321}

\bibitem{bradbury_jax_2018}
J.~Bradbury, R.~Frostig, P.~Hawkins, M.~J. Johnson, C.~Leary, D.~MacLaurin, G.~Necula, A.~Paszke, J.~VanderPlas, S.~Wanderman-Milne, Q.~Zhang, \href{http://github.com/jax-ml/jax}{{JAX}: composable transformations of {Python}+{NumPy} programs} (2018).
\newline\urlprefix\url{http://github.com/jax-ml/jax}

\bibitem{biewald_weights_2020}
L.~Biewald, C.~Van~Pelt, \href{https://wandb.ai/site/}{Weights \& {Biases}: {The} {AI} {Developer} {Platform}} (2020).
\newline\urlprefix\url{https://wandb.ai/site/}

\bibitem{hunter_matplotlib_2007}
J.~D. Hunter, \href{https://ieeexplore.ieee.org/document/4160265}{Matplotlib: {A} {2D} {Graphics} {Environment}}, Computing in Science \& Engineering 9~(3) (2007) 90--95, conference Name: Computing in Science \& Engineering.
\newblock \href {https://doi.org/10.1109/MCSE.2007.55} {\path{doi:10.1109/MCSE.2007.55}}.
\newline\urlprefix\url{https://ieeexplore.ieee.org/document/4160265}

\bibitem{burgers_mathematical_1948}
J.~M. Burgers, \href{https://www.sciencedirect.com/science/article/pii/S0065215608701005}{A {Mathematical} {Model} {Illustrating} the {Theory} of {Turbulence}}, in: R.~Von~Mises, T.~Von~Kármán (Eds.), Advances in {Applied} {Mechanics}, Vol.~1, Elsevier, 1948, pp. 171--199.
\newblock \href {https://doi.org/10.1016/S0065-2156(08)70100-5} {\path{doi:10.1016/S0065-2156(08)70100-5}}.
\newline\urlprefix\url{https://www.sciencedirect.com/science/article/pii/S0065215608701005}

\bibitem{rider_reconstructing_1998}
W.~J. Rider, D.~B. Kothe, \href{https://www.sciencedirect.com/science/article/pii/S002199919895906X}{Reconstructing {Volume} {Tracking}}, Journal of Computational Physics 141~(2) (1998) 112--152.
\newblock \href {https://doi.org/10.1006/jcph.1998.5906} {\path{doi:10.1006/jcph.1998.5906}}.
\newline\urlprefix\url{https://www.sciencedirect.com/science/article/pii/S002199919895906X}

\bibitem{virtanen_scipy_2020}
P.~Virtanen, R.~Gommers, T.~E. Oliphant, M.~Haberland, T.~Reddy, D.~Cournapeau, E.~Burovski, P.~Peterson, W.~Weckesser, J.~Bright, S.~J. van~der Walt, M.~Brett, J.~Wilson, K.~J. Millman, N.~Mayorov, A.~R.~J. Nelson, E.~Jones, R.~Kern, E.~Larson, C.~J. Carey, I.~Polat, Y.~Feng, E.~W. Moore, J.~VanderPlas, D.~Laxalde, J.~Perktold, R.~Cimrman, I.~Henriksen, E.~A. Quintero, C.~R. Harris, A.~M. Archibald, A.~H. Ribeiro, F.~Pedregosa, P.~van Mulbregt, {SciPy 1.0 Contributors}, {SciPy} 1.0: {Fundamental} {Algorithms} for {Scientific} {Computing} in {Python}, Nature Methods 17 (2020) 261--272.
\newblock \href {https://doi.org/10.1038/s41592-019-0686-2} {\path{doi:10.1038/s41592-019-0686-2}}.

\bibitem{driscoll_chebfun_2014}
T.~A. Driscoll, N.~Hale, L.~N. Trefethen, Chebfun guide, pafnuty publications Edition, Oxford, 2014.

\bibitem{mathias_augmenting_2022}
M.~S. Mathias, W.~P. de~Almeida, M.~R. de~Barros, J.~F. Coelho, L.~P. de~Freitas, F.~M. Moreno, C.~F.~D. Netto, F.~G. Cozman, A.~H.~R. Costa, E.~A. Tannuri, E.~S. Gomi, M.~Dottori, \href{http://arxiv.org/abs/2301.07824}{Augmenting a {Physics}-{Informed} {Neural} {Network} for the {2D} {Burgers} {Equation} by {Addition} of {Solution} {Data} {Points}}, arXiv:2301.07824 [physics] (2022).
\newblock \href {https://doi.org/10.1007/978-3-031-21689-3_28} {\path{doi:10.1007/978-3-031-21689-3_28}}.
\newline\urlprefix\url{http://arxiv.org/abs/2301.07824}

\end{thebibliography}

\appendix

\section{Burgers' results}
\label{app:burgers}
This section contains the results of the proposed PINN framework to the tested 1D and 2D Burgers' equations. This section also highlights the importance of different improvements based on ablation studies.

\subsection{Burgers' 1D}
\label{burgers1d}
The 1D Burgers' equation is a fundamental PDE that shares similarities with fluid flow problems. Thanks to its advection and diffusion terms, the PDE can model fluid dynamics behaviours such as shock formation and non-linear wave propagation. This benchmark provides us with insight into the PINN model's capability to capture nonlinear dynamics and handle the stability issues associated with high gradients and discontinuities. The 1D Burgers' equation, along with the initial and the Dirichlet boundary condition, is written as:
\begin{equation}
\begin{aligned}
    \frac{\partial u}{\partial t} + u\frac{\partial u}{\partial x} &= \nu\frac{\partial^2u}{\partial x^2},\quad t\in [0, T], \, x \in [-L,L],\\
    u(x=-L,t) &= u(x=L,t) = 0,\\
    u(x,t=0) &= 0
    \label{eq:burgers1d_pde}
\end{aligned}
\end{equation}
where \(u=u(x,t)\) represents the velocity field as a function of space and time and \(\nu\) is the kinematic viscosity. On the left side of the equation, the first term represents the time derivative. The second term is the advection term, describing how the velocity field transports itself through the domain and is responsible for the nonlinearity of the equation. On the right side, the diffusive term, controlled by the viscosity \(\nu\) smooths out the sharp gradients.
In our case, we set \(L=1\), \(T=1\), \(\nu=0.002\), and the characteristic velocity \(U=1\) giving us a high Reynolds number \(Re=1000\), making the problem challenging. We compare our PINN model with a reference solution generated using the Chebfun package in MATLAB \cite{driscoll_chebfun_2014}.  The reference solution is based on a spatial resolution of 201 points in time and 512 points in space, respectively.

To evaluate the efficiency of the features outlined in the methodology section \ref{sec:Improved_PINNs}, we conducted an ablation study. This study systematically disabled specific settings during the training phase to analyze their impact on the accuracy of the PINN model's predictions.  For consistency, we utilized an MLP architecture for all configurations as a first step. The \textit{PirateNet} model, however, utilizes the complete set of enhancements and employs the \textit{PirateNet} architecture instead of the MLP. The hyperparameter settings for the \textit{Plain}, \textit{Default}, and \textit{PirateNet} models are detailed in Table \ref{tab:burgers1d_hp}. Except for the architecture and the ablated hyperparameters, all other hyperparameters were kept constant.

\begin{figure}[!h]
    \centering
    \begin{subfigure}{1.0\textwidth}
        \centering
        \includegraphics[width=1.0\linewidth]{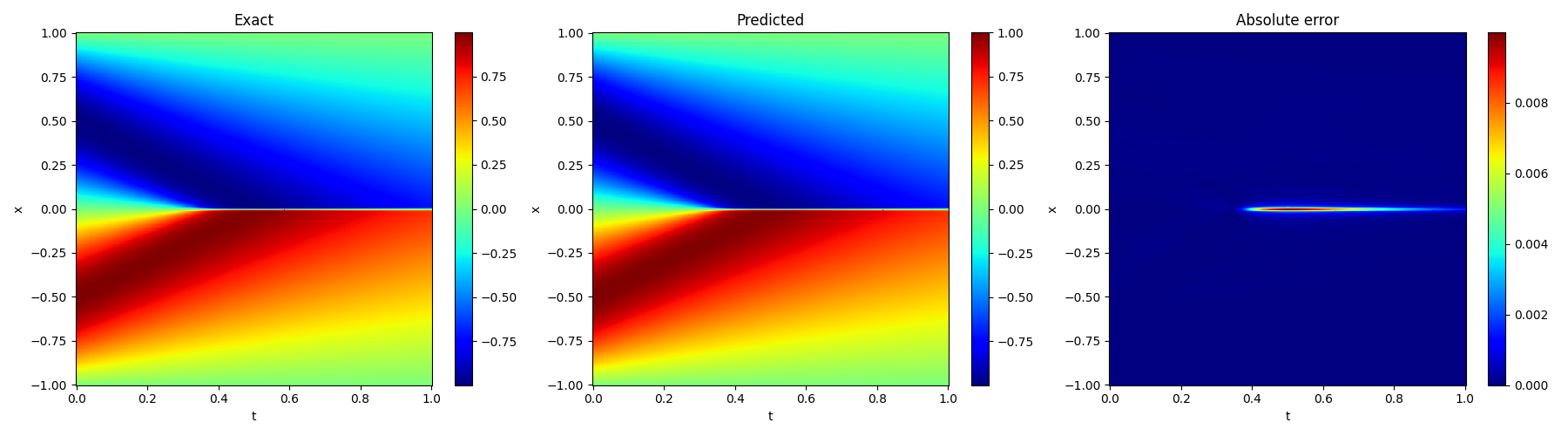}
        \caption{Burgers' 1D: Comparison of the \textit{PirateNet} prediction with the reference solution for the complete spatio-temporal grid.}
        \label{fig:burgers1d_piratenet_fullgrid}
    \end{subfigure}
    % Second subfigure
    \begin{subfigure}{1.0\textwidth}
        \centering
        \includegraphics[width=1.0\linewidth]{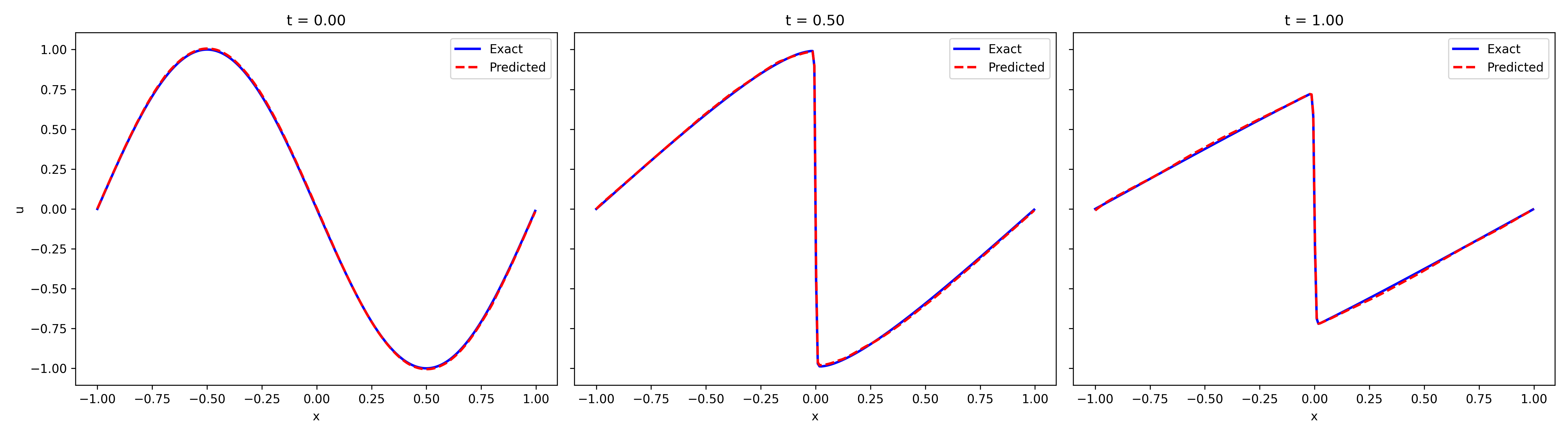}
        \caption{Burgers' 1D: Prediction and reference solution at time steps \(t =[0,0.5,1]\)s.}
        \label{fig:burgers1d_piratenet_timesteps}
    \end{subfigure}
    % Main figure caption
    \caption{Burgers' 1D: Comparison of the \textit{PirateNet} prediction with the reference solution.}
    \label{fig:burgers1d_piratenet}
\end{figure}

From Table~\ref{tab:burgers1d_ablation}, we can see that \textit{PirateNet} performs better than the other models, with a relative $L^2$-error of \(0.070\%\), whereas the worst by far is the \textit{Plain} PINN model with relative $L^2$-error of \(0.496\%\). We notice that the results of the best models are very close, showing that some features have less influence than others for the Burgers' 1D problem. However, it is suggested to keep all these features activated as they generally improve performance when solving PDEs with PINNs \cite{wang_experts_2023}. This is confirmed by the fact that the \textit{Default} configuration is very close to the other best-performing configurations with  \(L^2_{rel}=0.073\%\). The graphical solution for PINNs with the PirateNet architecture is shown in Figure \ref{fig:burgers1d_piratenet}. We can see from the absolute error plot that most of the error is concentrated at the point of shock formation. This can explain why most configurations have very similar \(L^2_{rel}\) errors. Most of the problem is easy to solve, making the harder area of the problem very small, resulting in a smaller impact on the \(L^2_{rel}\). The difference in model performances will be more obvious for other problems.

\begin{table}[ht]
    \centering
    \renewcommand{\arraystretch}{1.2}
    \setlength{\tabcolsep}{12pt}
    \small
    \begin{tabular}{l|c}
        \hline
        \textbf{Configuration name} & Relative \(L^2\)-error  (\%)\\
        \hline
         PirateNet & \textbf{0.070} \\
          No Causal Training & 0.072 \\
          Default & 0.073 \\
           No RWF & 0.073 \\
               No Fourier Feature & 0.093 \\
                No Grad Norm & 0.151 \\
        Plain & 0.496 \\
        \hline
    \end{tabular}
    \caption{Ablation study results for the 1D Burgers' equation: Relative \(L^2\) error comparison across configurations}
    \label{tab:burgers1d_ablation}
\end{table}

\subsection{Burgers' 2D}
\label{burgers2d}
The next benchmark problem is the 2D Burgers problem. It offers the same insight as its 1D counterpart but with an added spatial dimension as input and an added velocity field as output. The equations are written as:
\begin{equation}
    \frac{\partial u}{\partial t} + u\frac{\partial u}{\partial x} + v\frac{\partial u}{\partial y} = \nu\left(\frac{\partial^2u}{\partial x^2} + \frac{\partial^2u}{\partial y^2}\right),\quad t \in [0,T],\quad x,y \in [0,L],\\
    \label{eq:burgers2d_pde_u}
\end{equation}
\begin{equation}
    \frac{\partial v}{\partial t} + u\frac{\partial v}{\partial x} + v\frac{\partial v}{\partial y} = \nu\left(\frac{\partial^2v}{\partial x^2} + \frac{\partial^2v}{\partial y^2}\right),\quad t \in [0,T],\quad x,y \in [0,L],\\
    \label{eq:burgers2d_pde_v}
\end{equation}
The initial condition and the Dirichlet boundary conditions are given by:
\begin{equation}
    u(x,y, t=0)=\sin(2\pi x)\sin(2\pi y),
    \label{eq:burgers2d_ic_u}
\end{equation}
\begin{equation}
    v(x,y, t=0)=\sin(\pi x)\sin(\pi y),
    \label{eq:burgers2d_ic_v}
\end{equation}
\begin{equation}
    u(x,y, t)=0\quad \text{for}\quad (x,y) \in \partial\Omega,
    \label{eq:burgers2d_bc_u}
\end{equation}
\begin{equation}
    v(x,y, t)=0\quad \text{for}\quad (x,y) \in \partial\Omega,
    \label{eq:burgers2d_bc_v}
\end{equation}
In our case, we chose \(T = 0.5\), \(L = 1\), \(\nu=\frac{0.015}{\pi}\), which gives us a Reynolds number \(\text{Re} = 237\). The resolution in \(t\), \(x\) and \(y\) are all set to 101. The dataset used as a reference solution was generated using a finite differences method \cite{mathias_augmenting_2022}.

Similar to the Burgers' 1D case, we performed an ablation study to compare the \textit{PirateNet} architecture with other configurations under the same basic hyperparameters which can be found in Table \ref{tab:burgers2d_hp} for \textit{Plain}, \textit{Default}, and \textit{PirateNet}.

\begin{figure}
    \centering
    % Subfigure (a)
    \begin{subfigure}[b]{\linewidth}
        \centering
        \includegraphics[width=0.7\linewidth]{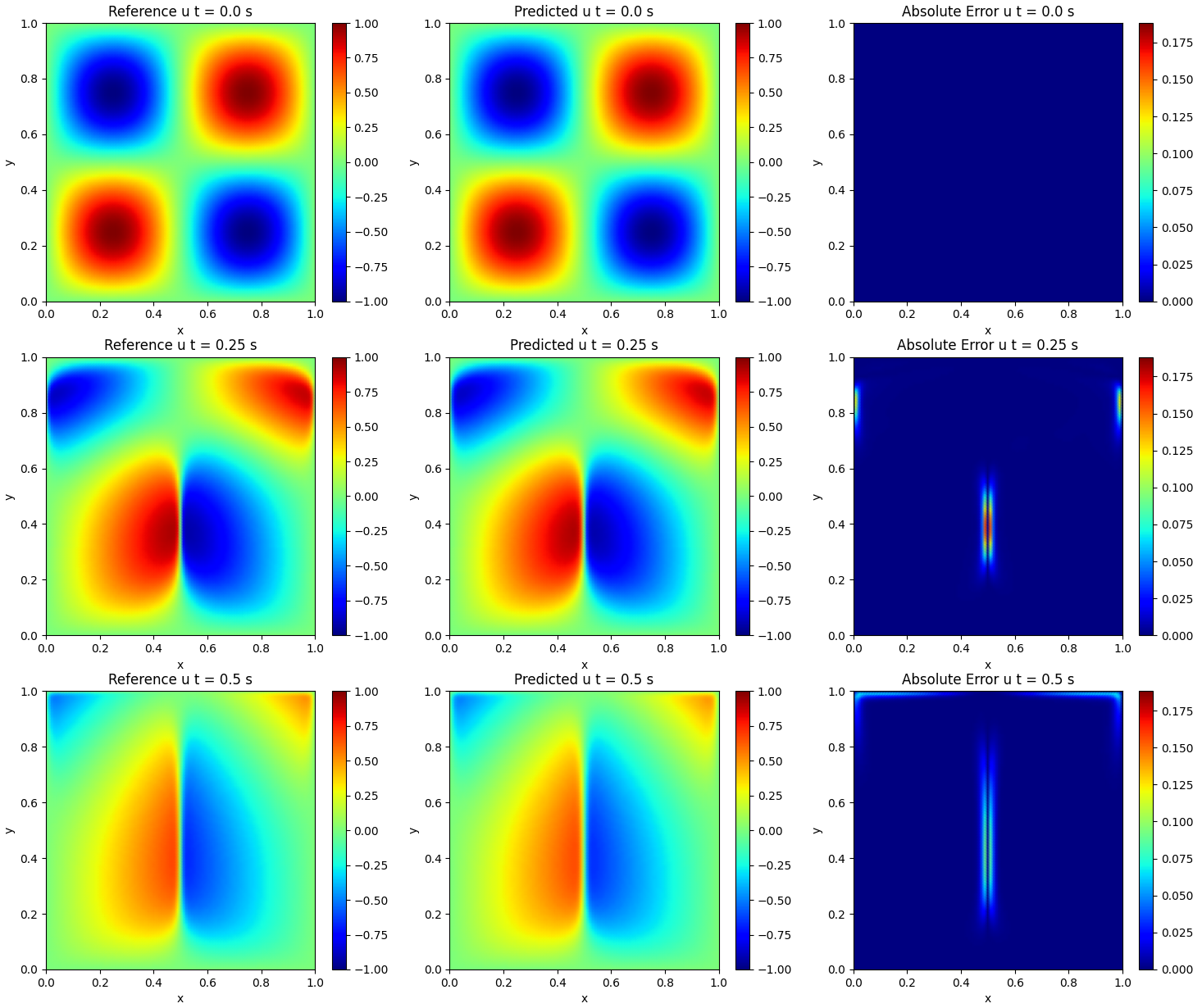}
        \caption{\(u\) velocity fields at \(t=[0,0.25,0.5]\)s}
        \label{fig:burgers2d_u}
    \end{subfigure}
    
    % Subfigure (b)
    \begin{subfigure}[b]{\linewidth}
        \centering
        \includegraphics[width=0.7\linewidth]{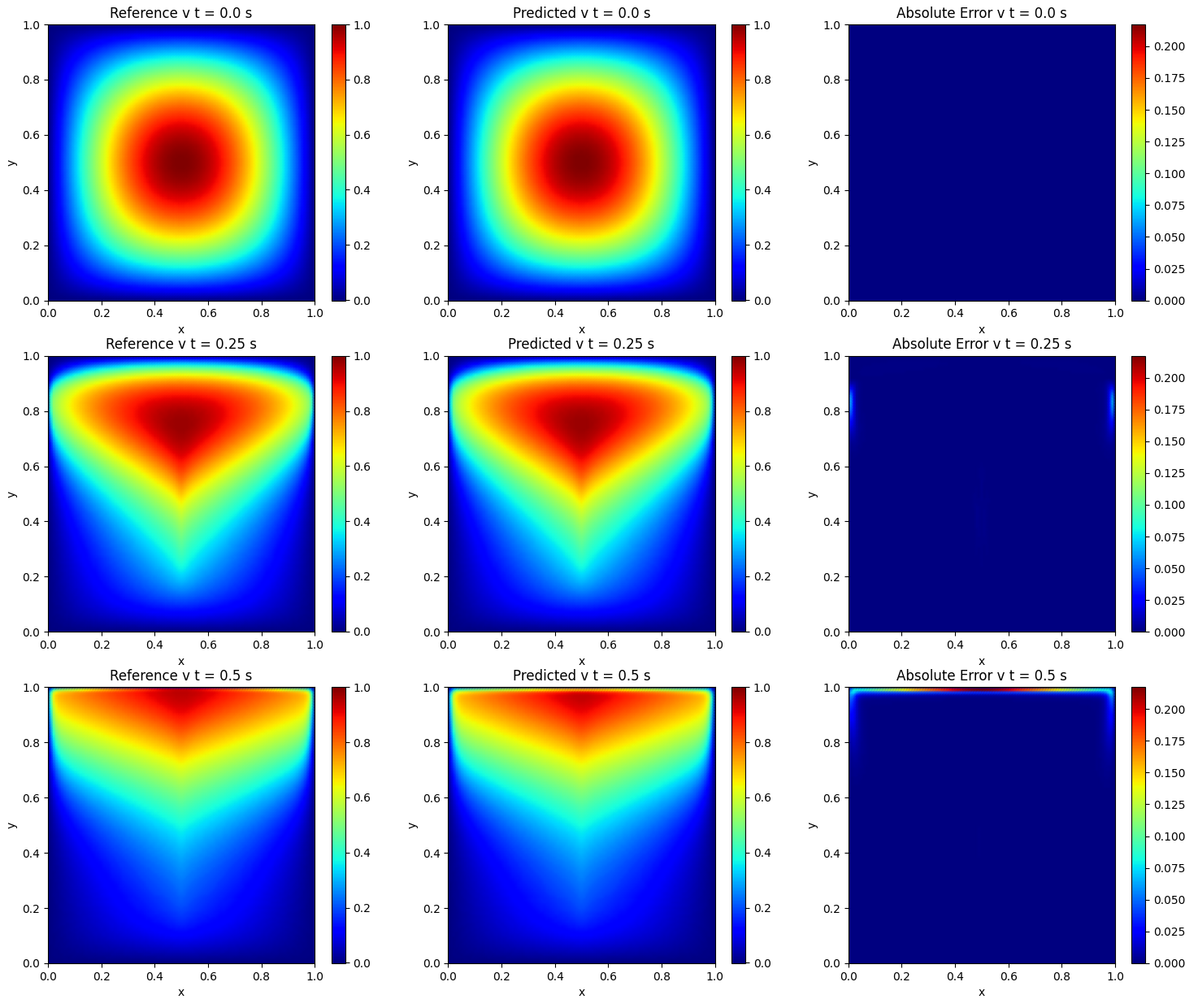}
        \caption{\(v\) velocity fields at \(t=[0,0.25,0.5]\)s}
        \label{fig:burgers2d_v}
    \end{subfigure}

    \caption{Reference, predicted and absolute error of \(u\), \(v\) velocity fields at different time steps with the \textit{PirateNet} configuration.}
    \label{fig:burgers2d_uv}
\end{figure}

Table \ref{tab:burgers2d_ablation} shows the \(L^2\) error for both outputted velocity fields \(u\) and \(v\), along with a norm of the errors \(L^2_{norm}=\sqrt{(L^2_{u})^2+(L^2_{v})^2}\). We can again see that \textit{PirateNet} performs better than the other configurations, with an \(L^2_{norm}=2.52\%\). A graphical representation of the reference and predicted velocity fields at different times is shown in Figure \ref{fig:burgers2d_uv} for the \textit{PirateNet}. Similar to Burgers' 1D, we notice that errors are very close to one another, which is expected because Burgers' 1D and Burgers' 2D are governed by the same type of equations. Again, most of the error is also near the wave front.

\begin{figure}[!h]
    \centering
    \includegraphics[width=1\linewidth]{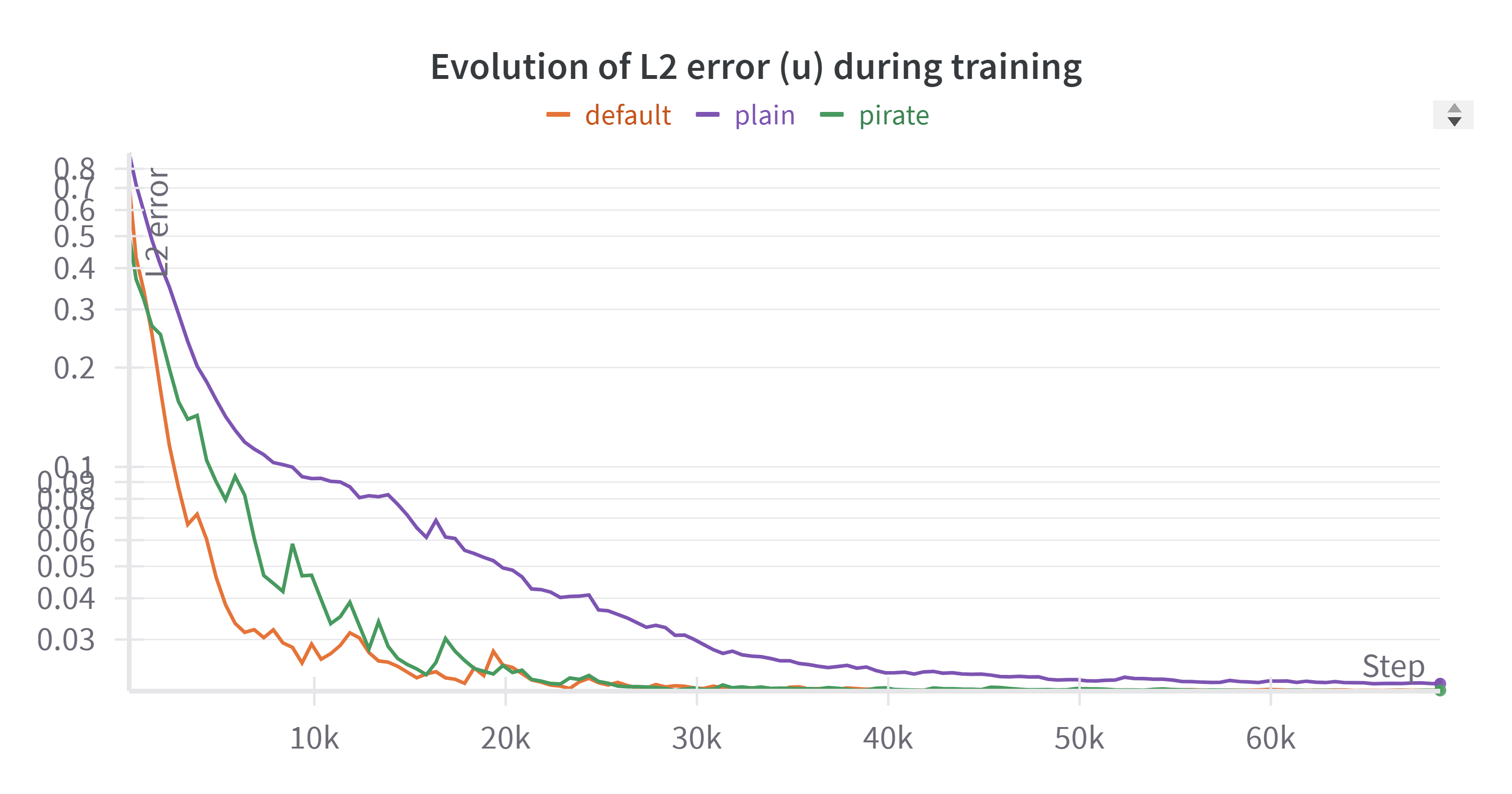}
    \caption{Evolution of \(L^2\) error during training for \(u\) and \(v\) velocity fields}
    \label{fig:burgers2d_l2_evolution}
\end{figure}

\begin{table}
    \centering
    \resizebox{\textwidth}{!}{%
        \begin{tabular}{c|cccc}
            \hline
             Configuration name& \(L_{2u}\) error (\%)& \(L_{2v}\) error (\%)&\(L_{2}\) norm (\%) &Run time (min)\\
             \hline
             Plain&  2.20& 1.46& 2.64&\textbf{57}\\
             Default&  2.11& 1.40& 2.53&128\\
             No Fourier feature&  2.12& 1.40& 2.54&100\\
             No RWF&  \textbf{2.10}& 1.40& 2.53&108\\
             No grad norm&  2.11& 1.40& 2.53&127\\
             No causal training&  \textbf{2.10}& \textbf{1.39}& \textbf{2.52}&148\\
             No modified MLP& 2.11& 1.40& \textbf{2.52}&61\\
             PirateNet& \textbf{ 2.10}& \textbf{1.39}& \textbf{2.52}&168\\
        \end{tabular}%
    }    
    \caption{Burgers' 2D: Ablation study.}
    \label{tab:burgers2d_ablation}
\end{table}
It is also worth noting the impact of certain features on computational load. Although run times can vary due to the state of the hardware and computational nodes of the compute cluster, there is a significant increase in run time when comparing the \textit{Plain} configuration with most other configurations with added features. The \textit{PirateNet} takes close to three times the time to train at 168 minutes, while the \textit{Default} takes more than twice the time at 128 minutes, compared to 57 minutes for the \textit{Plain} configuration. However, it is worth noting that this run time is for the complete 70k training steps. We can see from graph \ref{fig:burgers2d_l2_evolution} that \textit{PirateNet} and \textit{Default} converge much faster at around 25k steps compared to \textit{Plain} at around 40k.

Another aspect we tested was how a time-marching scheme could impact the results. 
Therefore, an S2S training scheme was implemented and tested for the \textit{PirateNet} configuration. As shown in Table \ref{tab:burgers2d_s2s_impact}, we can see that the S2S scheme with the \textit{PirateNet} architecture performs the best with an \(L^2_{norm}=2.49\%\), which highlights the sequential nature of the problem.
\begin{table}
    \centering
    %\resizebox{\textwidth}{!}{%
        \begin{tabular}{c|ccc}
            \hline
             Configuration name& \(L^2_{u}\) error (\%)& \(L^2_{v}\) error (\%)&\(L^{2}\) norm (\%) \\
             \hline
             PirateNet& 2.10& 1.39&2.52
    \\
             PirateNet S2S& \textbf{2.08}& \textbf{1.37}&\textbf{2.49}
    \\
        \end{tabular}
    %}
    \caption{Burgers' 2D: Impact of sequence-to-sequence training scheme}
    \label{tab:burgers2d_s2s_impact}
\end{table}

\section{Hyperparameters}
\label{hyperparameters}
The following appendix contains the hyperparameters used for training the models presented in the results section.

\begin{table}[ht]
    \centering
    \small
    \renewcommand{\arraystretch}{1}
    \setlength{\tabcolsep}{6pt}
    \resizebox{\textwidth}{!}{%
        \begin{tabular}{llccc}
        \hline
        \textbf{Category} & \textbf{Parameter} & \textbf{\textit{Plain}} & \textbf{\textit{Default}} & \textbf{\textit{PirateNet}}\\
        \hline
        \multirow{6}{*}{\textbf{Architecture}} 
        & Architecture & MLP & MLP & PirateNet\\
        & Number of layers & 4 & 4 & 4 \\
        & Layer size & 256 & 256 & 256 \\
        & Activation & Tanh & Tanh & Tanh \\
        & Fourier feature scale & -& 1.0 & 1.0 \\
        & RWF (\(\mu, \sigma\)) & -& \((0.5, 0.1)\) & \((0.5, 0.1)\) \\
        \hline
        \multirow{6}{*}{\textbf{Training}}& Optimizer & Adam & Adam & Adam \\
        & Training steps & 80,000 & 80,000 & 80,000 \\
        & Batch size & 4,096 & 4,096 & 4,096 \\
        & Learning rate & 0.001 & 0.001 & 0.001 \\
        & Decay steps & 2,000 & 2,000 & 2,000 \\
        & S2S windows & 1 & 1 & 1 \\
        \hline
        \multirow{4}{*}{\textbf{Weighting}} 
        & Scheme & Grad norm & Grad norm & Grad norm \\
        & Initial weights (\(\lambda_{ic}, \lambda_{bc}, \lambda_{r}\)) & (1, 1, 1) & (1, 1, 1) & (1, 1, 1) \\
        & Causal tolerance & -& 1.0 & 1.0 \\
        & Number of chunks & -& 32 & 32 \\
        \hline
        \end{tabular}
    }
    \caption{Burgers' 1D: Hyperparameters for the \textit{Default} configuration.}
    \label{tab:burgers1d_hp}
\end{table}

\begin{table}[ht]
    \centering
    \small
    \renewcommand{\arraystretch}{1}
    \setlength{\tabcolsep}{6pt}
    \resizebox{\textwidth}{!}{%
        \begin{tabular}{llccc}
        \hline
        \textbf{Category}&  \textbf{Parameter} & \textbf{\textit{Plain}}& \textbf{\textit{Default}} & \textbf{\textit{PirateNet}} \\
        \hline
        \multirow{6}{*}{\textbf{Architecture}}& Architecture & MLP & Modified MLP & PirateNet \\
        & Number of layers & 4 & 4 & 4 \\
        & Layer size & 256 & 256 & 256 \\
        & Activation & Gelu & Gelu & Gelu \\
        & Fourier feature scale & 1.0 & 1.0 & 1.0 \\
        & RWF (\(\mu, \sigma\))& (0.5, 0.1) & (0.5, 0.1) & (0.5, 0.1) \\
        \hline
        \multirow{6}{*}{\textbf{Training}}& Optimizer & Adam & Adam & Adam \\
        & Training steps per window& 70,000 & 70,000 & 70,000 \\
        & Batch size & 4,096 & 4,096 & 4,096 \\
        & Learning rate & 0.001 & 0.001 & 0.001 \\
        & Decay steps & 2,000 & 2,000 & 2,000 \\
        & S2S windows & 1 & 1 & 1 \\
        \hline
        \multirow{4}{*}{\textbf{Weighting}}& Scheme & Grad norm & Grad norm & Grad norm \\
        & Initial weights (\(\lambda_{ic}, \lambda_{bc}, \lambda_{r}\))& (10, 1, 1) & (10, 1, 1) & (10, 1, 1) \\
        & Causal tolerance & 1.0 & 1.0 & 1.0 \\
        & Number of chunks & 32 & 32 & 32 \\
        \end{tabular}
    }
    \caption{Burgers' 2D: Hyperparameters for the \textit{Default} configuration}
    \label{tab:burgers2d_hp}
\end{table}

\begin{table}[ht]
    \centering
    \small
    \renewcommand{\arraystretch}{1.2}
    \setlength{\tabcolsep}{6pt}
    \resizebox{\textwidth}{!}{%
        \begin{tabular}{llcccc}
        \hline
        &  \textbf{Parameter} &  \textbf{\textit{Plain}}
        &\textbf{\textit{Default}}&  \textbf{\textit{PirateNet}}&\textbf{\textit{Sota}}\\
        \hline
        \multirow{6}{*}{\textbf{Architecture}} & Architecture &  MLP & Modified MLP & PirateNet & PirateNet \\
        & Number of layers & 3 & 3 & 3 & 8 \\
        & Layer size & 256 & 256 & 256 & 256 \\
        & Activation &  Gelu& Gelu& Gelu& Swish \\
        & Fourier feature scale & - & 1.0 & 1.0 & 1.0\\
        & RWF (\(\mu\), \(\sigma\)) & - & (0.5, 0.1) & (0.5, 0.1) & (1.0, 0.1) \\
        \hline
        \multirow{6}{*}{\textbf{Training}}& Optimizer & Adam & Adam & Adam & Adam \\
        & Training steps & 20,000 & 20,000 & 20,000 & 80,000\\
        & Batch size & 2048 & 2048 & 2048 & 2048 \\
        & Learning rate & 0.001 & 0.001 & 0.001 & 0.001 \\
        & Decay steps & 2,000 & 2,000 & 2,000 & 2,000 \\
        & S2S windows & 1& 1& 1& 1\\
        & Physics-Informed initialization & False & False & True & True \\
        \hline
        \multirow{4}{*}{\textbf{Weighting}}& Scheme & - & Grad norm & Grad norm & Grad norm \\
        & Initial weights (\(\lambda_{ic}, \lambda_{r}\)) & (1, 1) & (1, 1) & (1, 1) & (1, 1) \\
        & Causal tolerance & - & 1.0 & 1.0 & 1.0\\
        & Number of chunks & - & 32 & 32 & 32 \\
        \end{tabular}
    }
    \caption{Level set Zalesak's disk: Hyperparameters for the \textit{Plain}, \textit{Default}, \textit{PirateNet} and \textit{sota} configurations}
    \label{tab:ls_zalesak_hp}
\end{table}

\begin{table}[ht]
    \centering
    \small
    \renewcommand{\arraystretch}{1.2}
    \setlength{\tabcolsep}{6pt}
    \resizebox{\textwidth}{!}{%
        \begin{tabular}{llcccc}
        \hline
        &  \textbf{Parameter} &  \textbf{\textit{Plain}}
        &\textbf{\textit{Default}}&  \textbf{\textit{PirateNet}}&\textbf{\textit{Sota}}\\
        \hline
        \multirow{6}{*}{\textbf{Architecture}} & Architecture &  MLP & Modified MLP & PirateNet & PirateNet \\
        & Number of layers & 3 & 3 & 3 & 8 \\
        & Layer size & 256 & 256 & 256 & 256 \\
        & Activation &  Relu & Relu & Relu & Swish \\
        & Fourier feature scale & - & 1.0 & 1.0 & 2.0 \\
        & RWF (\(\mu\), \(\sigma\)) & - & (0.5, 0.1) & (0.5, 0.1) & (1.0, 0.1) \\
        \hline
        \multirow{6}{*}{\textbf{Training}}& Optimizer & Adam & Adam & Adam & Adam \\
        & Training steps & 20,000 & 20,000 & 20,000 & 20,000 \\
        & Batch size & 2048 & 2048 & 2048 & 2048 \\
        & Learning rate & 0.001 & 0.001 & 0.001 & 0.001 \\
        & Decay steps & 2,000 & 2,000 & 2,000 & 2,000 \\
        & S2S windows & 2 & 2 & 2 & 2 \\
        & Physics-Informed initialization & False & False & True & True \\
        \hline
        \multirow{4}{*}{\textbf{Weighting}}& Scheme & - & Grad norm & Grad norm & Grad norm \\
        & Initial weights (\(\lambda_{ic}, \lambda_{r}\)) & (1, 1) & (1, 1) & (1, 1) & (1, 1) \\
        & Causal tolerance & - & 1.0 & 1.0 & 1.5 \\
        & Number of chunks & - & 32 & 32 & 32 \\
        \end{tabular}
    }
    \caption{Level set time-reversed vortex: Hyperparameters for the \textit{Plain}, \textit{Default}, \textit{PirateNet} and \textit{Sota} configurations}
    \label{tab:ls_vortex_hp}
\end{table}

\begin{table}[ht]
    \centering
    \small
    \renewcommand{\arraystretch}{1.2}
    \setlength{\tabcolsep}{6pt}
    \resizebox{\textwidth}{!}{%
        \begin{tabular}{llc}
        \hline
        &  \textbf{Parameter} &\textbf{\textit{Default}}\\
        \hline
        \multirow{6}{*}{\textbf{Architecture}} & Architecture & Modified MLP \\
        & Number of layers & 6\\
        & Layer size & 256 \\
        & Activation & Tanh\\
        & Fourier feature scale & 1.0 \\
        & RWF (\(\mu\), \(\sigma\)) & (1.0, 0.1)\\
        \hline
        \multirow{6}{*}{\textbf{Training}}& Optimizer & Adam \\
        & Training steps per window& 20,000 (40,000 for first window only)\\
        & Batch size & 4096\\
        & Learning rate & 0.001 \\
        & Decay steps & 2,000 \\
        & S2S windows & 16\\
        & Physics-Informed initialization & True\\
        \hline
        \multirow{4}{*}{\textbf{Weighting}}& Scheme & Grad norm \\
        & Initial weights (\(\lambda_{ic},\lambda_{bc}, \lambda_{r}\))& (1, 1, 1)\\
        & Causal tolerance & 1.0 \\
        & Number of chunks & 32 \\
        \end{tabular}
    }
    \caption{Coupled level set-NS: Hyperparameters}
    \label{tab:ls_coupled_hp}
\end{table}

\end{document}